%**start of header
%\input fiat
%**end of header
\newcount\mgnf\newcount\tipi\newcount\tipoformule
\newcount\aux\newcount\piepagina\newcount\xdata
\mgnf=0
\aux=1           %1 produce aux
\tipoformule=0   %0 usa aux; 1 no (usa i simboli dati)
\piepagina=1     %0 =data e #par.#pag; 1=data e #pag; 2=#pag
\xdata=0         %0 data del giorno, 1 data fissa da \Di:
\def\Di{}

\ifnum\mgnf=1 \aux=0 \tipoformule =1 \piepagina=1 \xdata=1\fi
\newcount\bibl
%\bibl= ?               % 0= rif [XXX], 1= rif. numerici
\ifnum\mgnf=0\bibl=0\else\bibl=1\fi
\bibl=0

% Per poter cambiare a piacimento il formato dei riferimenti
% bibliografici in <nome>.tex:
%
% 1: citare nella forma esemplificata da \ref{B}{2}{20}}
%    ove XXX e' un simbolo per le iniziali e 2 distingue i lavori con
%    le stesse iniziali,  7 e' il numero SIMBOLICO del riferimento per XXX2.
%    Il numero 7 puo' essere ARBITRARIO e viene automaticamente
%    riaggiustato al momento della compilazione (vedi punto 4)
%
% 2: Se si sceglie \bibl=0 si cita nella forma [XXX2]; se si sceglie
%    \bibl=1 si cita nella forma [numero di ordine di prima citazione].
%
% 3: La bibliografia va scritta nella forma \def{\qqq}{<citazione>}
%    in ordine alfabetico per autore attribuendo un simbolo
%    qualsiasi al testo <citazione} (ad es \def\REFCIT{citazione} e
%    va seguita dal comando  \rif{XXX}{2}{\qqq}{0}
%    scritto all' inizio di una riga altrimenti bianca. L' ultimo {0}
%    e' un parametro inutile ed e' li solo provvisoriamente per memoria
%    (di tentativi in corso).
%
% 4: Per il riaggiustamento automatico (in ordine di citazione)
%    occorre compilare (ottenendo oltre ai soliti la scheda ausiliaria
%    ref.b) e poi si deve eseguire il
%    programma rif <nome> che (usando ref.b) produce fin.tex e <nome>.tex
%    con i riferimenti giusti
%    in \bibl=1 e la si ricompila e stampa. La scheda iniziale <nome>.tex
%    diventa <nome>.old.
\ifnum\bibl=0
\def\ref#1#2#3{[#1#2]\write8{#1@#2}}
\def\rif#1#2#3#4{\item{[#1#2]} #3}
\fi

\ifnum\bibl=1
\openout8=ref.b
\def\ref#1#2#3{[#3]\write8{#1@#2}}
\def\rif#1#2#3#4{}

\fi

\def\9#1{\ifnum\aux=1#1\else\relax\fi}
\ifnum\piepagina=0 \footline={\rlap{\hbox{\copy200}\
$\st[\number\pageno]$}\hss\tenrm \foglio\hss}\fi \ifnum\piepagina=1
\footline={\rlap{\hbox{\copy200}} \hss\tenrm \folio\hss}\fi
\ifnum\piepagina=2\footline{\hss\tenrm\folio\hss}\fi

\ifnum\mgnf=0 \magnification=\magstep0
\hsize=13.5truecm\vsize=22.5truecm \parindent=4.pt\fi
\ifnum\mgnf=1 \magnification=\magstep1
\hsize=16.0truecm\vsize=22.5truecm\baselineskip24pt\vglue5.0truecm
\overfullrule=0pt \parindent=4.pt\fi

\let\a=\alpha\let\b=\beta \let\g=\gamma \let\d=\delta
\let\e=\varepsilon  \let\h=\eta
\let\th=\vartheta \let\l=\lambda \let\m=\mu \let\n=\nu
\let\x=\xi \let\p=\pi \let\r=\rho \let\s=\sigma 
 \let\f=\varphi\let\ch=\chi \let\ps=\psi \let\o=\omega
 \let\G=\Gamma \let\D=\Delta \let\Th=\Theta
\let\L=\Lambda   \let\F=\Phi
 \let\O=\Omega 
{\count255=\time\divide\count255 by 60 \xdef\oramin{\number\count255}
\multiply\count255 by-60\advance\count255 by\time
\xdef\oramin{\oramin:\ifnum\count255<10 0\fi\the\count255}}
\def\ora{\oramin }

%\Di e' definito all' inizio e ridefinito qui
\ifnum\xdata=0
\def\data{\number\day/\ifcase\month\or gennaio \or
febbraio \or marzo \or aprile \or maggio \or giugno \or luglio \or
agosto \or settembre \or ottobre \or novembre \or dicembre
\fi/\number\year;\ \ora}
\def\Di{\number\day\kern2mm\ifcase\month\or gennaio \or
febbraio \or marzo \or aprile \or maggio \or giugno \or luglio \or
agosto \or settembre \or ottobre \or novembre \or dicembre
\fi\kern0.1mm\number\year}
\else
\def\data{\Di}
\fi

\setbox200\hbox{$\scriptscriptstyle \data $}
\newcount\pgn \pgn=1
\def\foglio{\number\numsec:\number\pgn
\global\advance\pgn by 1} \def\foglioa{A\number\numsec:\number\pgn
\global\advance\pgn by 1}
\global\newcount\numsec\global\newcount\numfor \global\newcount\numfig
\gdef\profonditastruttura{\dp\strutbox}
\def\senondefinito#1{\expandafter\ifx\csname#1\endcsname\relax}
\def\SIA #1,#2,#3 {\senondefinito{#1#2} \expandafter\xdef\csname
#1#2\endcsname{#3} \else \write16{???? ma #1,#2 e' gia' stato definito
!!!!} \fi} \def\etichetta(#1){(\veroparagrafo.\veraformula) \SIA
e,#1,(\veroparagrafo.\veraformula) \global\advance\numfor by 1
\9{\write15{\string\FU (#1){\equ(#1)}}} \9{ \write16{ EQ \equ(#1) == #1
}}} \def \FU(#1)#2{\SIA fu,#1,#2 }
\def\etichettaa(#1){(A\veroparagrafo.\veraformula) \SIA
e,#1,(A\veroparagrafo.\veraformula) \global\advance\numfor by 1
\9{\write15{\string\FU (#1){\equ(#1)}}} \9{ \write16{ EQ \equ(#1) == #1
}}} \def\getichetta(#1){Fig.  \verafigura \SIA e,#1,{\verafigura}
\global\advance\numfig by 1 \9{\write15{\string\FU (#1){\equ(#1)}}} \9{
\write16{ Fig.  \equ(#1) ha simbolo #1 }}} \newdimen\gwidth \def\BOZZA{
\def\alato(##1){ {\vtop to \profonditastruttura{\baselineskip
\profonditastruttura\vss
\rlap{\kern-\hsize\kern-1.2truecm{$\scriptstyle##1$}}}}}
\def\galato(##1){ \gwidth=\hsize \divide\gwidth by 2 {\vtop to
\profonditastruttura{\baselineskip \profonditastruttura\vss
\rlap{\kern-\gwidth\kern-1.2truecm{$\scriptstyle##1$}}}}} }
\def\alato(#1){} \def\galato(#1){}
\def\veroparagrafo{\number\numsec}\def\veraformula{\number\numfor}
\def\verafigura{\number\numfig}
\def\geq(#1){\getichetta(#1)\galato(#1)}
\def\Eq(#1){\eqno{\etichetta(#1)\alato(#1)}}
\def\eq(#1){\etichetta(#1)\alato(#1)}
\def\Eqa(#1){\eqno{\etichettaa(#1)\alato(#1)}}
\def\eqa(#1){\etichettaa(#1)\alato(#1)}
\def\eqv(#1){\senondefinito{fu#1}$\clubsuit$#1\write16{No translation
for #1} \else\csname fu#1\endcsname\fi}
\def\equ(#1){\senondefinito{e#1}\eqv(#1)\else\csname e#1\endcsname\fi}
\openin13=#1.aux \ifeof13 \relax \else \input #1.aux \closein13\fi
\openin14=\jobname.aux \ifeof14 \relax \else \input \jobname.aux
\closein14 \fi \9{\openout15=\jobname.aux} \newskip\ttglue

%\font\dodicirm=cmr12\font\dodicibf=cmbx12\font\dodiciit=cmti12
%\font\titolo=cmbx12 scaled \magstep2
%\font\ottorm=cmr8\font\ottoi=cmmi8\font\ottosy=cmsy8
%\font\ottobf=cmbx8\font\ottott=cmtt8\font\ottosl=cmsl8\font\ottoit=cmti8
%\font\sixrm=cmr6\font\sixbf=cmbx6\font\sixi=cmmi6\font\sixsy=cmsy6

\font\titolone=cmbx12 scaled \magstep2

\font\ottorm=cmr8\font\ottoi=cmmi7\font\ottosy=cmsy7
\font\ottobf=cmbx7\font\ottott=cmtt8\font\ottosl=cmsl8\font\ottoit=cmti7
\font\sixrm=cmr6\font\sixbf=cmbx7\font\sixi=cmmi7\font\sixsy=cmsy7

\font\fiverm=cmr5\font\fivesy=cmsy5\font\fivei=cmmi5\font\fivebf=cmbx5
\def\ottopunti{\def\rm{\fam0\ottorm}\textfont0=\ottorm%
\scriptfont0=\sixrm\scriptscriptfont0=\fiverm\textfont1=\ottoi%
\scriptfont1=\sixi\scriptscriptfont1=\fivei\textfont2=\ottosy%
\scriptfont2=\sixsy\scriptscriptfont2=\fivesy\textfont3=\tenex%
\scriptfont3=\tenex\scriptscriptfont3=\tenex\textfont\itfam=\ottoit%
\def\it{\fam\itfam\ottoit}\textfont\slfam=\ottosl%
\def\sl{\fam\slfam\ottosl}\textfont\ttfam=\ottott%
\def\tt{\fam\ttfam\ottott}\textfont\bffam=\ottobf%
\scriptfont\bffam=\sixbf\scriptscriptfont\bffam=\fivebf%
\def\bf{\fam\bffam\ottobf}\tt\ttglue=.5em plus.25em minus.15em%
\setbox\strutbox=\hbox{\vrule height7pt depth2pt width0pt}%
\normalbaselineskip=9pt\let\sc=\sixrm\normalbaselines\rm}

\catcode`@=11
\def\footnote#1{\edef\@sf{\spacefactor\the\spacefactor}#1\@sf
\insert\footins\bgroup\ottopunti\interlinepenalty100\let\par=\endgraf
\leftskip=0pt \rightskip=0pt \splittopskip=10pt plus 1pt minus 1pt
\floatingpenalty=20000
\smallskip\item{#1}\bgroup\strut\aftergroup\@foot\let\next}
\skip\footins=12pt plus 2pt minus 4pt\dimen\footins=30pc\catcode`@=12
\let\nota=\ottopunti

%% Grafica
\newdimen\xshift \newdimen\xwidth \newdimen\yshift

\def\ins#1#2#3{\vbox to0pt{\kern-#2 \hbox{\kern#1
#3}\vss}\nointerlineskip}

\def\eqfig#1#2#3#4#5{ \par\xwidth=#1
\xshift=\hsize \advance\xshift by-\xwidth \divide\xshift by 2
\yshift=#2 \divide\yshift by 2 \line{\hglue\xshift \vbox to #2{\vfil #3
\includegraphics{#4.ps} }\hfill\raise\yshift\hbox{#5}}} 

\def\8{\write13}

\def\figini#1{\catcode`\%=12\catcode`\{=12\catcode`\}=12
\catcode`\<=1\catcode`\>=2\openout13=#1.ps}

\def\figfin{\closeout13\catcode`\%=14\catcode`\{=1
\catcode`\}=2\catcode`\<=12\catcode`\>=12}

%%%%%%%%%%%%%%%%%%%%%%

\def\V#1{{\,\underline#1\,}}
\def\T#1{#1\kern-4pt\lower9pt\hbox{$\widetilde{}$}\kern4pt{}}
\let\dpr=\partial \let\io=\infty\let\ig=\int
\def\fra#1#2{{#1\over#2}}\def\media#1{\langle{#1}\rangle}\let\0=\noindent
\def\guida{\leaders\hbox to 1em{\hss.\hss}\hfill}
\def\tende#1{\ \vtop{\ialign{##\crcr\rightarrowfill\crcr
\noalign{\kern-1pt\nointerlineskip} \hglue3.pt${\scriptstyle%
#1}$\hglue3.pt\crcr}}\,} \def\otto{\
{\kern-1.truept\leftarrow\kern-5.truept\to\kern-1.truept}\ }

\def\tto{{\Rightarrow}}
\def\pagina{\vfill\eject}

\def\st{\scriptscriptstyle}
\def\*{\vskip0.3truecm}

\def\lis#1{{\overline #1}}\def\eg{\hbox{\it e.g.\ }}

\def\ie{\hbox{\it i.e.\ }}

\def\fiat{{}}
\def\\{\hfill\break} \def\={{ \; \equiv \; }}

\def\annota#1{\footnote{${{}^{\bf#1}}$}}
\ifnum\aux=1\BOZZA\else\relax\fi
\ifnum\tipoformule=1\let\Eq=\eqno\def\eq{}\let\Eqa=\eqno\def\eqa{}
\def\equ{{}}\fi
\def\defi{\,{\buildrel def \over =}\,}

\def\1{\ifnum\mgnf=0\pagina\else\relax\fi}
\def\W#1{#1_{\kern-3pt\lower6.6truept\hbox to 1.1truemm
{$\widetilde{}$\hfill}}\kern0pt}

\def\AA{{\cal A}}\def\EE{{\cal E}}\def\xx{{\V x}}
\def\GG{{\cal G}}\def\cfr{{\it c.f.r.\ }}

\def\kk{{\V k}}\def\uu{{\V u}}
\def\VV#1{{\underline
#1}_{\kern-3pt$\lower7pt\hbox{$\widetilde{}$}}\kern3pt\,}
\def\Tr{{\,{\rm Tr}\,}}

\fiat

\font\msytw=msbm10 scaled\magstep1

\def\RRR{\hbox{\msytw R}}

\def\yy{{\underline y}}     \def\dd{{\underline d}}
\def\ttu{{\underline t}}    \def\uu{{\underline u}}

\0\centerline{\titolone 
Large deviations in rarefied quantum gases}
\*

\centerline{\it G. Gallavotti, J.L. Lebowitz, V. Mastropietro}
\centerline{\it Rutgers}
\*
\0{\bf Abstract:} {\it 
The probability of observing a large deviation (LD) in the number of
particles in a region $\Lambda$ in a dilute quantum gas contained in a
much larger region $V$ is shown to decay as $\exp[-|\Lambda|\Delta
F\,]$, where $|\L|$ is the volume of $\Lambda$ and $\Delta F$ is the 
change in the appropriate free energy density, the same as in
classical systems.  However, in contrast with the classical case,
where this formula holds at all temperatures and chemical potentials
our proof is restricted to rarefied gases, both for the typical and
observed density, at least for Bose or Fermi systems. The case of
Boltzmann statistics with a bounded repulsive potential can be treated
at all temperatures and densities. Fermions on a lattice in any
dimension, or in the continuum in one dimension, can be treated at all
densities and temperatures if the interaction is small enough
(depending on density and temperature), provided one assumes periodic
boundary conditions.}
\*

\0{\bf 1. Introduction.}
\numsec=1\numfor=1\*

We study the probability distribution of the number of particles in a
box $\L$ for a quantum system in a region $V \supset
\Lambda$ described by a grand-canonical ensemble with reciprocal
temperature $\beta$ and chemical potential $\mu$. We are primarily
interested in the case where the volume of $V$ is much larger than
that of
$\Lambda$, $|V|>>|\Lambda|$, i.e. in properties remaining 
valid in the thermodynamic limit $V \to \RRR^d$ with $\mu,\beta$
fixed and $|\L|/|V|\to0$. The particles interact with a pair potential
$v(x-y)$ so the  Hamiltonian of the system is given by
$$H_V=\sum_{i=1}
(-\fra12\D_{x_i}-\m)+ \sum_{i<j} v(x_i-x_j), \quad
x_i \in V\Eq(1.1)$$
where the Laplace operator is considered with Dirichlet boundary
conditions
on the walls of $V$, which we assume for simplicity to be a cube with
side
length $L$ (centered at the origin), or with periodic boundary
conditions
in the Fermionic case at small $v$ considered in sec. 7.  The analysis
in
sections 3--6 carries over to the cases of Neumann, periodic or mixed
boundary conditions and for most (reasonable) shapes of the containers
$V$
and $\L$. This does not imply that effects of the boundary conditions
in
quantum systems are well understood: it just means that in sections
3--6 we
do not consider values of $\mu$ and $T$ where phase transitions are
possible. In fact the cases considered in  sec. 7. make essential use
of the
periodicity in the boundary conditions.

The potential  $v(x)$ is assumed to vanish  for $|x| > D$ and to
be bounded, smooth and stable. The particles will be
assumed to obey either Boltzmann, Bose or Fermi statistics.
Hard core interactions  can also be treated, leading to results
analogous
to those of Sections  3--6 but the Fermionic cases in Section 7 would 
require new
ideas in the presence of hard cores. For simplicity we do not consider
this type of interaction here.

The  number of particles in a region $\L$ which, again for the sake of
simplicity, we take to be a cube of side length $\ell$ centered at the
origin is given by $N_\L(\V x)=\sum_i
\ch_\L(x_i)$ where $\V x=(x_1,\ldots)$ denotes a configuration of
particles in $V$ and $\ch_\L$ is the indicator function of the box
$\L$ ($\ch_\L(x)=1$ if $x\in\L$ and  $\ch_\L(x)=0$ otherwise). The
dimension of the space will be taken $d=3$ but our analysis holds in
any dimension.  The probability of finding
exactly $n$ particles in the box $\L$ is then given by

$$\Pi(n)= \fra{\Tr \d_n e^{-\b H_V}}{\Tr  e^{-\b H_V}}\Eq(1.2)$$
where $\d_n(\V x)=1$ if $N_\L(\V x)=n$ and $0$ otherwise.

We note that formally the only difference in $\Pi(n)$ between quantum
and classical systems, is the nature of the probability distribution
of configurations ${\V x}$ in $V$.  For a classical system this is
given by a Gibbs measure while for a quantum system it has to be
computed via the density matrix.

The pressure $P$, Helmholtz free energy $f$ and expected density
$\rho$ at temperature $\b^{-1}$ and chemical potential $\m$ are given
as usual by

$$
\b P(\b,\m)=\lim_{L\to\io} L^{-d}\log \Tr e^{-\b H_V},\qquad
\b f(\b,\r)= \inf_\mu\{\b\m\r-\b P(\b,\m)\}\Eq(1.3)$$
We shall restrict our analysis to values of $\b,\m$ where the above
functions are differentiable and the extremum is achieved at a single
point $\m=\m(\beta,\r)$ so that  the expected particle density in $V$

$$\r=\fra{\dpr}{\dpr \m} P(\b,\m),\qquad \m=\fra{\dpr}{\dpr \r}
f(\b,\r)\Eq(1.4)$$
We call the region in which this differentiability holds the ``no
phase transitions region''. This is a larger region than the region
$\AA(\GG)$ where the functions $P,\r,f$ have been proved by Ginibre,
see the review [Gi71], to be analytic in $z=e^{\b \m}$:   a disk in
the $z$--plane with radius
$R(\b)>0$ (an estimate for $R(\b)$ is quoted later, see \equ(3.4)).
The region
$\AA(\GG)$ is also a region where upper and lower bounds on the
derivatives ${\partial P \over\partial \rho},\fra{\dpr f}{\dpr \m}$
can be established: 

 Let 
 $$\eqalign {\D F(\b,\r,\r_0) & \defi 
\,f(\b,\r)-f(\b,\r_0)-
\fra{\dpr f(\b,\r)}{\dpr\r}\Big|_{\r=\r_0}\,(\r-\r_0)\cr ~~~ & = [f(\b,\r) -
\mu(\b,\r_0)\r] - [f(\b,\r_0) - \mu(\b,\r_0)\r_0]\cr},\Eq(1.5)$$
$\Delta F$ is the difference between the  Helmholtz free energy at
density $\r$ and its linear
extrapolation from $\rho_0$ (which is positive where (1.4) holds if
$\rho \ne \rho_0$).   We shall prove that the probability of finding a
density $\rho$ in $\Lambda$, when the expected density in $V$ is
$\rho_0$,
is given asymptotically by

$$\Pi(n \simeq \rho \ell^d)\sim e^{-\b \D F(\b,\r,\r_0)\,
\ell^d}\Eq(1.6)$$
for large $\ell$, provided $\r-\r_0$ is small enough. More
mathematically this is stated as follows:
\*

\0{\it Let  $\b,\m$ be fixed in the analyticity region $\AA(\GG)$ and
let $\r_0$ be the corresponding density (so that
$\r_0=\r(\beta,\mu)$).
If the side $\ell$ of the box $\L$ tends to infinity and,
correspondingly, the container side  also tends to infinity so that
$L/\ell\to\io$ then

$$\lim_{\ell\to\io} {\ell^{-d}}\log\sum_{\tilde \r\in[a,b]} 
\Pi(\tilde\r\ell^d)=
\max_{\tilde \r\in{[a,b]}} \, -\b\,\D F(\b,\tilde \r,\r_0)\Eq(1.7)$$
This holds for all statistics if the interval $[a,b]$ is
contained in an interval  $[-\d_0(\b,\m),$  $\d_0(\b,\m)]$ centered at
$\r_0$ with $\d_0(\b,\m)>0$ small enough (estimated in \equ(3.4)).}
\*

In Section 5 we obtain \equ(1.7) by proving its finite $\ell$
version (which contains various finite size corrections).
We  then  prove, in 6, a similar result for Boltzmann Statistics with 
arbitrary  $(\b,\m)$,  $\b>0$  assuming a
bounded repulsive (\ie positive) interaction $v$: this extends the
validity of \equ(1.6) far beyond the analyticity region $\AA(\GG)$. We
can also treat Bose statistics in a region somewhat larger than
$\AA(\GG)$ but still under the very restrictive assumption that
$e^{\b\m+2\b B}<1$ where $B$ is the stability constant of the
interaction potential $v$, \ie $\sum_{i<j=1}^n v(x_i-x_j)>-B n$ for
all $(x_1,\ldots,x_n)$.   Our method does not seem extendable to more
general  Bosonic
systems in spite of the strong results of Park,
sec. 5 of [Pa85].  
\*

A further extension in Section 7, using completely
different techniques (see [GM00] for a review), deals with
weakly interacting Fermi systems on a lattice: given $\b,\r_0$ the
above theorem holds if the interaction potential is small enough
(depending on density and temperature) and if {\it periodic boundary
conditions are assumed on the container $V$}. The results  can also be
extended to continuum systems in $d=1$.  They  hold
for values of $\mu$, where the method of Sec. 2-6 fails.
\*

The above restrictions always exclude the region in  $\b,\m$ where
genuine
quantum phase transitions may occur (like superfluidity or
superconductivity).  Substantial further work appears necessary to
deal
with this regime despite the fact that the ideal Fermi and Bose gas
(no
interactions) can be treated completely by other methods, see [LLS99].
\*

\0{\bf 2. The Ginibre representation.}
\numsec=2\numfor=1\*

The key technical ingredient in our analysis is the Ginibre
representation of the quantum partition functions, [Gi71]. We describe
here only that part of the formalism which we need to derive
our results.

Let $\O=(x,\o)$ be a Brownian path starting at $x$ at $t=0$ and
returning to $x$ at a later time $b$: this is a continuous function
$t\to \o(t)\in \RRR^d$ defined for $t\in[0,b]$. The time length $b$
will
be fixed as $b=\b$ in the case of Boltzmann statistics but it will
take the values $b=\b,2\b,3\b,\ldots$ in the case of Bose or Fermi
statistics.

Functions of $\O$ will be integrated with a measure $d\O$ which is
defined as $d\O=dx\cdot P_{xx}(d\o)$ in terms of the
{\it conditional Wiener measures} $P_{xy}^b(d\o)$ (see [Gi71]
p. 343);\annota{1}{This is simply defined to be the measure on the
paths starting at $x$ and ending at $y$ in a  time $b$ formally given
by
$P^b_{xy}(d\o)\defi P_x(d\o)\d(\o(b)-y)$ where $P_x(d\o)$ is the usual
Wiener distribution}

$$\ig F(\O)d\O\defi \sum_{j=1}^{*} \fra{(-1)^{(j-1)\s} }{j}\ig dx\ig
P_{xx}^{j\b}(d\o) F(\O)\Eq(2.1)$$
where, see [Gi71] p.361, the statistics are distinguished by the upper
limit $*$ and the sign exponent $\s$ as
$$
\cases{*=1& Boltzmann statistics\cr
*=\io,\, \s=0& Bose statistics\cr
*=\io,\,\s=+1 & Fermi statistics\cr}\Eq(2.2)$$
and the integration is over all  $x\in V$ and $\o(t)\in V$ for all
$t$.

For $\O_k=(x_k,\o_k)$ such that $\o_k$ has
time length $j_k\b$ we imagine that $\o_k$ consists of $j_k$ strings,
each of time length $\b$ with the $i$--th string denoted by
$\o_{k,i}(t)=\o_k(i\b+t)$, $t\in[0,\b]$.
 We can consider each bit of string of length $\beta$ as a
``particle''
and the collection  $\V\O=(\O_1,\ldots,\O_n)$ as ``trajectory
configurations'' or as ``configurations of particles delocalized by
quantum indeterminacy''
 The ``energy'' of a
configuration $\V\Omega$ is  then defined as

$$U(\V\O)=\fra1{2\,\b} \sum_{(k,i)\ne (k',i')}\ig_0^\b
v(\o_{k,i}(t)-\o_{k',i'}(t))\, dt\Eq(2.3)$$
\*
which is consistent with the intuitive delocalization
interpretation above  in which the ``number of particles'' in the
configuration
$\V\O$  is simply $j_1+\ldots+j_n$.  It is convenient to
introduce two notions of number of particles of $\V\O$ inside $\L$
as

$$%\eqalign{
%&
N_\L(\V\O)= \sum_{(k,i)}\ch_\L(\o_{k,i}(0)),\qquad
%\cr&
{\tilde N}_\L(\V\O)=\b^{-1}\sum_{(k,i)} \ig_0^\b
\ch_\L(\o_{k,i}(t))\,dt
%\cr}
\Eq(2.4)$$
and note that $N_\L(\V\O)$ is an integer while ${\tilde N}_\L(\V\O)$
is generally not, except that ${\tilde N}_V({\V\Omega}) =
N_V({\V\Omega})$ is always so.
%\end

The remarkable representation for the grand canonical partition
function
(due to Ginibre) is,

$$Z(\b,\m;V)\defi
\Tr e^{-\beta H_V} =\sum_{n=0}^\io \ig_{N_V(\V\O)=n} z^n \, e^{-\b
U(\V\O)}\,\fra{d\V\O}{n!}
\Eq(2.5)$$
where $\V\O=(\O_1,\ldots,\O_n)$ and $z\defi e^{\b \m}$ defines the
{\it activity}.  (2.5) makes the quantum partition function look like
the classical partition function of a gas of closed contours of length
roughly $\sqrt{\b}$ for all values of $z$ in the Boltzmann statistics
and for $z e^{2\b B}<1$ in the Bose and Fermi case; $B$ is a
stability constant defined by $U(x_1,\dots,x_n)\ge-B n, \forall n$.

Remarkably, the quantum reduced density matrices also admit a natural
``classical'' representation in terms of the above contours but we
shall not need such a representation here, see [Gi71].
\*

\0{\bf 3. The cluster expansion for rarefied gases.}
\numsec=3\numfor=1\*

One of the important consequences of the above classical
representation of a quantum system is that it immediately allows us to
take advantage of the techniques developed to study Mayer and
virial expansions of classical gases. The basic remark is that it is
``easy'' to take the logarithm of a classical partition function. By
simple algebraic considerations the logarithm is obtained as a
formal power series in $z$ and the difficult part is to show the
convergence of such an expansion.

The formal power series expression of the grand canonical pressure $\b
P(\b,\m)$, defined in (1.3) is obtained in the form:

$$\log Z(\beta,\mu,V) = \beta P(\beta, \mu;V) =\sum_{n=0}^\io
\ig_{N_V(\V\O)=n} z^n \,\F^T(\V\O)\,\fra{d\V\O}{n!},\Eq(3.1)$$
The r.h.s. of \equ(3.1) still looks like a partition function with
$\F^T(\V\O)$ replacing $e^{-\b U(\V\O)}$, see eq. (4.15) in
[Gi71]. However, {\it unlike } the functions $e^{-\b U(\V\O)}$, the
functions $\F^T(\V\O)$ have the ``{\it cluster property}'' of
``decay'' at infinity, i.e.\ if ${\V \Omega} = ({\V \O}',
{\V \O}'')$ consists of trajectories
$\V\O'=(\O_1',\ldots,\O_{n'}')$, $\V\O''=(\O_1'',\ldots\O_{n''}'')$
with
$(\V\O',\V\O'') \defi
(\O'_1,\ldots,\O'_{n'},\O_1'',\ldots\O_{n''}'')$,
then

$$|\F^T({\V \Omega}^\prime, \V\O'')|=0 \qquad {\rm
if}\ d(\V\O^\prime,\V\O'')>D\Eq(3.2)$$
where  $d(A,B)$ is the distance of two sets $A,B$ and the
trajectories are considered here as the  union of the sets of points
they occupy as time varies. This is an elementary property that
follows from the explicit expression for the
functions $\F^T$ in terms of ``Mayer graphs'', see eq. (4.3) in [Gi71]
(see also [GMM71], eq. (4.2) and (4.3) at p. 176, for a similar
simpler case).

The functions $\F^T(\V\O)$ are like the functions $e^{-\b U(\V\O)}$
{\it translation invariant} in the sense that they have the same value
for $\V\O$ and for a translate of $\V\O$ as long as the two contour
configurations are inside $V$: this is a property that will be used to
guarantee the existence of the limit as $L\to\io$.

Furthermore trajectories that are long give a small contribution to
\equ(3.1)  at small fugacities because, see  eq. (4.39) in[Gi71],

$$\ig_{N(\V\Th)=p} |d\,\V\Th|\, |\Phi^T(\V\O,\V\Th)|< (2^{-1} e^{-\b
w})^{q-1} R^{-(p+q-1)}\Eq(3.3)$$
 Here, $q = N(\V\O)$ and $w$ is the maximum of $v$, $|d\,\V\Th|$ means
that we use the Bosonic measure (which
maximizes the integral) whether the system verifies Bose, Fermi
or Boltzmann statistics, $N(\V\Th)$ is the number of
elementary trajectories composing the configuration $\V\Th$ and
$$R=R(\b) =2^{-1}\,e \exp{-\b\,w \,-2\b\, B-\b \,l(\b)^{-d} e^{\b B}
C\,\ig
|v(x)|dx}\Eq(3.4)$$
In (3.4) $l(\b)=\sqrt{2\p\b}$ is the ``thermal length'' and
$C=\sum_{j=1}^\io 2^{-j} j^{-d/2}$, see [Gi71], eq. (3.16) (here we
made a special choice for the parameter $\x$ in [Gi71], namely
$\x=2^{-1} e^{-\b\, w}$). Since $w<\io$ hard cores cannot be
considered. The bound in \equ(3.3) holds uniformly in $V$ and it can
be considerably improved in the case of Boltzmann statistics, see
[Gi71] eq. (3.15).

Note the asymptotic values of $R(\b)$: $R(\io)=0$ always and
$R(0)=0$  if $d>2$ while $R(0)=r_0>0$ if $d\le 2$.

We shall further need the following remark: suppose that the partition
function  defined in (2.5) is altered by inserting a factor $\f(\V\O)$
which has the property of factoring:

$$\f(\V\O', \V\O'')=\f(\V\O')\f(\V\O'')\Eq(3.5)$$
for all $\V\O', \V\O''$ then (remarkably)
$$\eqalign{
&\log Z_\phi(\b,\m;V)\defi\log
\sum_{n=0}^\io \ig_{N_V(\V\O)=n} z^n \, \f(\V\O)\, \fra{d\V\O}{n!}
e^{-\b
U(\V\O)}=\cr
&=\sum_{n=0}^\io \ig_{N_V(\V\O)=n} z^n\, \f(\V\O)\F^T(\V\O)\,
\fra{d\V\O}{n!}
\cr}\Eq(3.6)$$

Consequently if $|z| \max_\O |\f(\O)|<R$ the series in
\equ(3.6) converges uniformly in the size $L$ of the container and the
sums in the r.h.s. of \equ(3.1) and \equ(3.6) are therefore convergent
representations of the logarithms of the partition functions \equ(2.5)
and \equ(3.6) that define them. The bound \equ(3.3)
 goes back to
the theory of the Kirkwood--Salsburg equations and of the Mayer and
virial expansions, see [Ru69], [Gi71].
%\end

\*
\0{\bf 4. Laplace transform of the probability.}
\numsec=4\numfor=1\*

Turning back to the large deviations problem it is natural to look for
properties of the Laplace transform (generating function) $\G(\l)$ of
the probability distribution $\Pi(n)$,  defined in (1.2)

$$\G(\l)\defi\sum_{n=0}^\io e^{\b\l n} \Pi(n)\= \fra{\Tr e^{\b\l N_\L}
e^{-\b H_V}} {\Tr e^{-\b H_V}}\Eq(4.1)$$
This admits again a simple Brownian path representation:
$$\Gamma(\lambda) = \fra{\sum_{n=0}^\io \ig_{N_V(\V\O)=n} z^n
e^{\l\b N_\L(\V\O)} e^{-\b U(\V\O)}
\fra{d\V\O}{n!}}{\sum_{n=0}^\io \ig_{N_V(\V\O)=n} z^n e^{-\b U(\V\O)}
\fra{d\V\O}{n!}}\Eq(4.2)$$
Note that $N_\L(\V\O)$ defined in \equ(2.4) appears here. The similar
expression with $N_\L(\V\O)$ replaced by ${\tilde N}_\L(\V\O)$ is a
representation of the ratio:

$$\fra{\Tr e^{\b\l {N}_\L-\b H_V}}
{\Tr e^{-\b H_V}}\Eq(4.3)$$
which is related to \equ(4.1) as we shall see, but which is
interesting
in its own right.

The theory of  sections 2 and 3 applies to the expressions \equ(4.1),
\equ(4.3): in
fact, since the functions $\f(\V\O)=e^{\l \b N_\L(\V\O)}$ have the
factorization property \equ(3.5), eq. \equ(4.1) can be written as

$$\G(\l)=\exp\,\big(\sum_{n=0}^\io \ig_{\V\O\cap
\L\ne\emptyset\atop N_V(\V\O)=n} z^n \,(e^{\l\b
N_\L(\V\O)}-1)\,\F^T(\V\O)\,\fra{d\V\O}{n!}\big)\Eq(4.4)$$
 Let now $z$ be less than $R$ then (\cfr \equ(3.4)), if 
$$|z|\, |e^{\b\l}|<R\qquad {\rm or}\qquad
\l<\d(\b,\m)=\b^{-1}(\log R-\b\m)\Eq(4.5)$$
 the right side of (4.4) converges to a limit as $V\to\io$.  This
limit of $\Gamma(\lambda)$ is
uniform in $V$.
Clearly this is a key point that holds
because of the bounds \equ(3.3) and \equ(3.4). 
Furthermore the same convergence bound implies that the argument of
the exponential in \equ(4.4) can be differentiated term by term
with respect to $\l$ still yielding  convergent series.
%\end

Conclusion: the generating function \equ(4.1) is analytic in $\l$ in a
small interval $|\lambda| < \delta(\beta,\mu)$ around $0$.  In fact
one has
analyticity in a bigger region of $\l$ which contains the infinite
real semi-axis $\l<\d(\b,\m)$.

We define for each $\L$, $|\L| = l^d$,

$$\eqalign{
&\b P_\ell(\b,\m;\l)=\ell^{-d}
\sum_{n=0}^\io \ig_{\V\O\cap \L\ne\emptyset\atop N_V(\V\O)=n}
 z^n
(e^{\l\b N_\L(\V\O)}-1)\,\F^T(\V\O)
\fra{d\V\O}{n!}\cr
&\r_\ell(\b,\m;\l)=
{\ell^{-d}} \sum_{n=0}^\io \ig_{\V\O\cap \L\ne\emptyset\atop
N_V(\V\O)=n}
 z^n\,
e^{\l\b N_\L(\V\O)}N_\L(\V\O)\,\F^T(\V\O)
\fra{d\V\O}{n!}=\dpr_\l P_\ell(\b,\m;\l)\cr
}\Eq(4.6)$$
The convergence bounds in \equ(3.3) and the translation invariance of
the $\F^T$ functions then imply that the above expressions have limits
as $\ell\to\io$ provided that the size of the container $V$ is also
such that $L/\ell\to\io$.  (Indeed the integrals over $\V\O$ in (4.6)
can be written as the difference between integrals over completely
arbitrary configurations
$\V\O$  except for the restriction that they
contain one point (\eg $x_1$) inside $\L$ and integrals over
configurations that intersect the boundary.  The latter contribute a
quantity
of $o(\ell^{d})$ in the expression for $P$ and $o(1)$ in the
expression for the density).\annota{2}{In fact a closer analysis would
reveal that such terms can be estimated to have size ${O(\ell^{d-1})}$
and $O(\ell^{-1})$, \ie that they are boundary terms, see [Gi71] proof
of lemma 2.2, p.366.} The limits  of $\beta P_l$ and $\rho_l$ are
given by

$$\eqalign{
&\b (P(\b,\m+\l)-P(\b,\m))=
\sum_{n=0}^\io \ig^*_{N(\V\O)=n}
 z^n
(e^{\l\b  n}-1)\,\F^T(\V\O)\,
\fra{d\V\O}{n!}\cr
&\r(\b,\m+\l)=
 \sum_{n=1}^\io \ig^*_{N(\V\O)=n}
 z^n
e^{\l\b n} n\,\F^T(\V\O)
\fra{d\V\O}{n!}=\dpr_\l P(\b,\m+\l)\cr
}\Eq(4.7)$$
where the $*$ over the integral means that the point $x_1=\o_{1,1}(0)$
is not integrated over; it can be fixed at the origin.

The approach to the limits is uniform in the parameters $\b,\m$ in any
closed region contained in the analyticity domain $\AA(\GG)$. One
could also evaluate the finite $\ell$ corrections and show that if
$L/\ell\ge2$ (say) then their sizes are of order $\ell^{-1}$ and in
fact consist of two terms of respective orders $L^{-1}$ and
$\ell^{-1}$ (\ie quantities of the order of ``surface/volume'' coming
from boundary effects due to the boundaries of $V$ and of $\L$
respectively): however one would have to enter into the details of
Ginibre's work, so that we just note that \equ(3.3) and translation
invariance imply that the corrections go to zero as $\ell\to\io$.
\*
%\1

\0{\bf 5. Large deviations in the analyticity region \equ(4.5).}
\numsec=5\numfor=1\*

Given the above information, obtained from [Gi71], the derivation of
the large deviations results follows the standard path set up in the
classical theory of [MS67].

For $\b$ and $\m=\m_0$ we call, \cfr \equ(4.5),
$\r_\ell=\r_\ell(\b,\m_0;0)$ and denote the grand canonical averages
at chemical potential $\m$ by $\media{\cdot}_\m$. We estimate the
probability $\Pi$ that $N_\L> (\r_\ell+a) \ell^d$, for some small
$a>0$,
by
$$\Pi(N_\L> (\r_\ell +a) \ell^d)\le
\media{e^{\l\b\,(N_\L- (\r_\ell+a)
\ell^d)}}_{\m_0}=e^{\b\,\big(P_\ell(\b,\m_0;\l)-P_\ell(\b,\m_0;0)
-\l\,(\r_\ell+a)\big)\ell^d}\Eq(5.1)$$
for all $\l\ge0$. The limit of the coefficient of $\ell^d$ in the
r.h.s. of \equ(5.1) is, by \equ(4.7),
$\b\,\big((P(\b,\m_0+\l)-P(\b,\m_0)-\dpr_\m P(\b,\m_0) \l-\l a\big)$,
which holds for all $\l\ge0$. Minimizing over $\l$ we get (by
definition of free energy), that {\it if $a$ is small enough}
$\Pi(N_\L>(\r_\ell+a)\ell^d)\le e^{-\b a^2\ell^d/2\chi_++o(\ell^d)}$
where $\chi_+$ is the minimum compressibility (\ie of the second
derivative of $P$ with respect to $\m$) in the interval
$[\m_0-\d,\m_0+\d]$. This is within the radius of convergence of our
expansion, provided the value $\l\sim a/\chi(0)$, where the minimum is
achieved, is such that $|\l|<\d(\m_0,\b)$, \ie $|a|<
\d_0(\b,\m_0)\defi\chi_+ \d(\b,\m_0)$ so that the above estimates on
the
Laplace transform hold.

The case $a<0$ can be treated in a similar manner. Hence setting
$|a|<\d_0(\b,\m)$ we get
\*
(I) {\it Large deviation property 1:}
$$\Pi(|N_\L-\r\ell^d|>a\ell^d)\le e^{-\b a^2\ell^d/2\chi_++o(\ell^d)}
\Eq(5.2)$$
{\it for $a\in(0,\d_0(\b,\m_0))$.}
\*

The relation \equ(5.2) gives an upper bound on 
arbitrarily large fluctuations which, since the number of possible
values of $N$ between $-a\ell^d $ and
$a\ell^d$ is ``only'' $2a\ell^d$ and the total probability is
$1$ implies that inside any density interval of size
$|a|\le\d_0(\b,\m_0)$ centered at $\r_0$ there must be at least one
value of $N$ whose probability is $\ge O(\ell^{-d})$. We write this
result as
\*
(II) {\it Large deviation property 2:}

$$\Pi(N_\L\,{\buildrel a \over \sim}\, \r_0 \ell^d)\ge
O(a^{-1}\ell^{-d})
\Eq(5.3)$$
{\it where $N{\buildrel \e \over \sim} \r\ell^d$ denotes the event
$|N-\r_0\ell^d|<\e\ell^d$.}
\*

This shows that the numbers $\sim \r_0 \ell^d$ have a ``high
probability'', and is our second ``large deviations'' result.
\*

We now consider a density value $\r_0+a$ corresponding to a chemical
potential $\m_0+\l_a$, with $a\in (0,\d_0(\m_0,\b))$, and note that
the same argument can be applied to the probability distribution
generated by replacing $e^{-\beta H_V}$ in (1.2) by $e^{-\b \l_a
N_\L/2} e^{-\b
H_V}e^{-\b\l_a N_\L/2} $. Calling  $\tilde \Pi(N_\L)$ this
distribution
we therefore conclude that

$$\tilde \Pi(N_\L\,{\buildrel \e \over \sim}\,
(\r_0+a)\ell^d)\ge O(\e^{-1}
\ell^{-d}) e^{o(\ell^d)}\Eq(5.4)$$
Therefore, with the notation introduced after \equ(5.3),

$$\eqalign{
&\Pi(N_\L {\buildrel \e \over \sim} (\r_0+a)\ell^d)=\fra{\sum_{
N_\L\,{\buildrel \e \over \sim}\, (\r_0+a)\ell^d}
\Tr e^{-\b H_V }e^{\b \l_a\,N_\L}}
{\Tr e^{-\b H_V} e^{\b \l_a\,N_\L}}\cdot\cr
&\cdot\, e^{-\l_a \b
(\r_0+a)\ell^d}\,
\fra{\Tr e^{-\b H_V} e^{\b\l_a\,N_\L}}{\Tr e^{-\b H_V} }
\, e^{O(\e\ell^{d})}=\cr &=
\tilde \Pi(N_\L {\buildrel \e \over \sim} (\r_0+a)\ell^d)
e^{-\l_a \b (\r_0+a)\ell^d}
\fra{\Tr e^{-\b H_V} e^{\b\l_a\,N_\L}}{\Tr e^{-\b H_V}}\,
e^{O(\e\ell^{d})}\cr}\Eq(5.5)$$
where we have multiplied and divided by $\Tr e^{-\b H_V} e^{\b
\l_a\,N_\L}$ and introduced in the first tern in the l.h.s.
$e^{\b \l_a\,N_\L}$ compensating it by $e^{-\l_a \b(\r_0+a)\ell^d}
e^{O(\ell^d)}$ which is possible because of the restriction
$N_\L\,{\buildrel \e \over \sim}\, (\r_0+a)\ell^d$ in the sum.

The first term is between $1$ and $O(\e^{-1}\l^{-d} e^{-O(\e\ell^d)})$
by \equ(5.4) and the last ratio can be written, again as done above
and in
the previous section, as

$$\exp\,\big( \sum_{n=0}^\io \ig_{N(\V\O)=n} z^{N(\V\O)}\,
(e^{\b \l_a N_\L(\V\O)}-1)\,\fra{d\V\O}{n!}\big)\Eq(5.6)$$
where the integrals over $\V\O$ can be divided into regions with
$\V\O$ entirely outside the box $\L$, which cancel because
$N_\L(\V\O)\=0$, and into those partly inside and partly outside,
which
contribute a quantity of $o(\ell^{d}),$\annota{3}{Once more a closer
analysis  shows that such terms can be estimated to have size
$O(\ell^{d-1})$, \ie that they are boundary terms.} and those
entirely inside $\L$, which by the analysis in sec. 4 give $\ell^d\,\b
\,(P(\b,\m_0+\l_a) -P(\b,\m_0))=o(\ell^d)$.

Hence we see that
\*
(III) {\it Large deviations property 3:}

$$\eqalign{ &\Pi(N_\L\,{\buildrel \e \over \sim}\, (\r_0+a) \ell^d)=
e^{\ell^d\,\b\,\big(-\l_a (\r_0+a) +
P(\b,\m_0+\l_a)-P(\m_0)\big)}\,O(\e^{-1}\ell^{-d}
e^{O(\e\ell^d)})=\cr &= e^{-\b
\ell^d\, \big(f(\b,\r_0+a)-f(\r_0)-\dpr_\r
f(\b,\r_0)\,a\,\big)}\,O(\e^{-1}\ell^{-d}e^{O(\e\,\ell^d)})\cr}\Eq(5.7)$$
having used the relations $\b P(\b,\m_0)=\b\m_0\r_0-\b
f(\b,\r_0)$ if $\m_0=\dpr_\r\,f(\b,\r_0)$ (see [Ru69], eq. (4.17)).

The smoothness of the functions $f, P$ has been tacitly used in the
above discussion. This allows
us to take advantage of the arbitrariness of $\e$ (which we could even
choose to be an inverse power of $\ell$ with a small exponent) and to
interchange limits and maximizations so that \equ(5.7) implies the
result stated in  sec. 1. Indeed we deduce that in the interval
$[\r_0+a,\r_0+b]\ell^d$ there is a value of $N$ whose probability
multiplied by $e^{\D F(\,\b,\r_0+a)
\ell^d}$ is $\ge O(e^{(b-a)O(\ell^d)} \ell^{-d})$ times
the exponential in \equ(5.7) and $a,b$ are arbitrary in
$(0,\d_0(\b,\m_0))$. This is the main result claimed in Sec. 1.
\*

\0{\bf 6. Beyond the analyticity region.}
\numsec=6\numfor=1\*

A completely different method can be used to study systems
satisfying Boltzmann statistics at {\it arbitrary chemical potential}
or Bose gases {\it in the region $e^{\b\m+2\b B}<1$}. The method
does not apply to Fermi systems.
We illustrate it in the bounded positive interaction case: more
general cases can presumably be treated along similar lines but we
shall not attempt to do so here.

The method is very similar to the one used in classical statistical
mechanics by [La72], [Ol88].  This entails comparing $\Gamma(\lambda)$
in (4.1) with the partition function $Z(\beta,\mu+\lambda;\Lambda)$
obtained by replacing $H_V$ in the right side of (4.1) with
$H_\Lambda$.  More precisely we want to prove that $|\Lambda|^{-1}
\log \Gamma(\lambda) \to |\Lambda|^{-1} \log Z(\beta,
\mu+\lambda;\Lambda) = P(\beta, \mu + \lambda)$ which would give us
directly the desired result.  For a classical system this is done by
first
noting that $\Gamma(\lambda)$ is (for $L - l > D)$ nothing else than
$\int
Z(\beta,\mu + \lambda; \Lambda|\eta) d \nu(\eta)$.  Here $\eta$ is the
configuration of particles outside $\Lambda$, $d\nu(\eta)$ is the
induced
Gibbs measure and $Z(\beta,\mu+\lambda;\Lambda|\eta)$ is the partition
function (or normalization factor) for the grand canonical ensemble 
with chemical potential $\mu + \lambda$, in a box $\Lambda$ with
boundary
conditions $\eta$.  The proof of the LD formula is then basically a
version
of the proof of equivalence of ensembles, [Ru69].

To carry out a similar analysis for the quantum Boltzmann and Bose
case it
is necessary to use trajectories $\V\O$ instead of configurations.
 One simply considers
probabilities of events happening in a fictitious {\it gas of closed
paths}: the probability of a configuration $\V\O=(\O_1,\ldots,\O_n)$
of this gas is,

$$z^n e^{-\beta U(\V\O)} d\V\O/Z(\beta,\mu,V)\Eq(6.1)$$
with the notation of the previous sections.  We can
now write $\V\O = (\V\O_\Lambda,\V \O_{ext})$ in which
the contours in $\V\O_\Lambda$ are totally or partially inside $\L$
and those
in $\V\O_{ext}$ {\it do not have any points inside $\L$}. Then the
probability distribution in
\equ(6.1) induces conditional probabilities $W_\L(\V\O_\Lambda
|\,\V\O_{ext})$
verifying the appropriate ``{\it DLR equations}''

$$W_\L(\V\O\,|\,\V\O_{ext})=\fra{ z^n e^{\b
U(\V\O\,|\,\V\O_{ext})}\,d\V\O/n!}
{\tilde Z(\beta,\mu;\Lambda|\V\O_{ext})}
%  {\sum_{n'} \ig_{N_\L(\V\O')=n'}
%z^{n'} e^{-\b U(\V\O'\,|\,\V\O_{ext})} d\V\O'/{n'!}}
\Eq(6.2)$$
where $\tilde Z$ is the appropriate normalization.
(Fermions cannot be treated in this way because the integration
$d\V\O$
is {\it not positive}).

The main estimate on which the analysis will rely is that, in analogy
to the classical case, the contribution to the partition function from
trajectories which cross $\partial \Lambda$ is proportional to
$\partial \Lambda$.  More precisely there exists a function
$\th(\b,z)$ such that
$$\fra{1}{|\dpr\L|} \log \sum_m \ig_{N_V(\V\O)=m, \O_1\cap
\dpr\L\not=\emptyset}^{\widetilde{}}
z^m e^{-\b U(\V\O\,|\,\V\O')} d\V\O/{m!}<
\th(\b,z)<\io\Eq(6.3)$$
for all $\L\subseteq V$ and for all configurations $\V\O'$; the
tilde over the integral means that the contours in $\V\O$ must all
intersect $\L$. This is an estimate that can be proved without
restrictions in the case of a Boltzmann statistics; however in the
case of bosons it can only be proved under the condition $z e^{2\b
B}<1$; for a proof see [Gi71] proof of lemma 2.2, p.366.\annota{4}{In
an unpublished note by Lupini and one of us, (GG), this region was
slightly extended to cover cases with $z e^{2\b B}>1$ in systems with
hard core interactions.  This made use of the fact that hard cores
force the Brownian paths $\O$ to be ``quite extended'' (\ie $O(a)$ if
$a$ is the core radius) while they have average size $O(\sqrt\b)$:
therefore they have a low probability if $\b$ is small. Hence one can
take $z e^{2\b B}>1$ (by $O(e^{a^2/\b}$) provided $\b$ is small
enough. The analysis could be extended to cover also such cases which,
however, are conceptually not really different from the case
considered here.}

The probability $\Pi(N)$ that $N_\L({\V\O})=N$ can be estimated by
considering the probability of the same event conditioned on the
presence of an external configuration $\V\O_{ext}$: if the resulting
estimate coincides with \equ(1.6) up to a correction by a factor
$e^{O(\ell^{d-1})}$ {\it independently} of $\V\O_{ext}$ the proof of
\equ(1.7) will be achieved.
\*

\0{(IV) \it Large deviation property 4:}
\*

\0{\it Suppose that the interaction potential $v$
is positive and bounded. Let $\b,\m_0$ be in the region $e^{\b
\m_0+2\b B}<1$ in the Bose case or arbitrary in the Boltzmann case. We
also assume that there is no phase transition at $\b,\m_0$. Let $\r_0$
be the corresponding density (so that $\r_0=\r(\beta,\m_0)$), i.e.\
${\partial P(\beta, \mu) \over \partial \mu}$, is continuous at
$\mu_0$.  If the side $\ell$ of the box $\L$ tends to infinity and,
correspondingly, the container side $L$ also tends to infinity so that
$L/\ell\to\io$, then

$$\lim_{\ell\to\io} {\ell^{-d}}\log\sum_{\tilde \r\in[a,b]} 
\Pi(\tilde \r\ell^d)=
\max_{\tilde \r\in{[a,b]}} \, -\b\,\D F(\b,\tilde \r,\r_0)\Eq(6.4)$$
This holds for the Bose system provided the interval $[a,b]$ is
contained in an interval $(0,\r_{max}(\b))$ where $\r_{max}(\b)$ is
the maximal density that can be achieved as the chemical potential
varies compatibly with the restriction $e^{\b \m_0+2\b B}<1$.  (There
is no such restriction in the Boltzmann case).}
\*

The assumption of boundedness on the potential does not allow hard
cores. However the technique below also applies to the hard core case
with a tail which is not necessarily repulsive: to avoid
technicalities we do not discuss this case. Eliminating completely all
restrictions, \ie assuming just stability or possibly superstability
of the potential, seems to require new ideas.

To prove \equ(6.4) we fix the configuration $\V\O_{ext}$ outside
$\L$ and we look first for a lower bound for $\Pi_(n)$. This is
obtained by remarking that if $\V\O\subset\L$ then for configurations
$\V\O$ in which all trajectories are at least a distance $D$ from
$\partial \Lambda$, $|U(\V\O\,|\,\V\O_{ext})|=0$.

Hence, if $\L_D$ is the region in the interior of $\L$  separated from
the
boundary of $\dpr\L$ by a corridor of width $D$, the numerator of
the fraction (obtained from \equ(6.2))
expressing $\Pi(n)$ can be bounded below by $
\big(\ig_{\V\O\subset\L_D, n_\L(\O)=n} e^{\b\m_0 n}
e^{-U(\V\O)}d\V\O/n!\big)$ which is $e^{\b\m_0 n}$ times the {\it
canonical} partition function $Q^D_n$ for the system with
Dirichlet boundary conditions (\ie va\-ni\-shing boundary conditions)
on $\dpr\L_D$ and no external particles outside the box $\L_D$.

The denominator in \equ(6.2) is bounded above by $Z_{\L}^0(\m_0,\b)
e^{\th(\m_0,\b)|\dpr \L|}$ where $Z^0_{\L}$ is the grand
canonical partition function for the system with Dirichlet boundary
conditions on $\dpr\L$ and no external particles
outside the box $\L$; the second factor  bounds the contributions to
the integral from the contours of $\V\O$ that cross the boundary {\it
via} the bound \equ(6.3). Hence

$$\Pi_{\L}(N)\ge \fra{e^{\b\m_0 N} Q^D_N}{Z^0_{\L}(\m_0,\b)}
e^{-\th(\m_0,\b)\,|\dpr\L|}\Eq(6.5)$$
Likewise $Z^0_{\L} $ gives a lower bound to the denominator
of \equ(6.2) while an upper bound on the numerator is

$$\Big(\sum_{k=0}^N \ig_{N_\L(\V\O)=N-k}
e^{\b\m_0 (N-k)}\, e^{-\b U(\V\O)}\Big)
e^{\th(\b,\m_0)|\dpr\L|}\Eq(6.6)$$
and again we recognize that the integral gives the canonical partition
function for $N-k$ particles in $\dpr\L$ with Dirichlet (and no
outside
particles) boundary condition multiplied by $e^{\b\m_0(N-k)}$.
Therefore

$$\eqalignno{
&\Pi_{\L,N}\le e^{\th|\dpr\L|}\,N\,\max_{k\le N}
\fra{e^{\b\m_0 (N-k)} Q_{N-k}(\b)}{Z^0_{\L_D}(\b,\m_0)}=\cr 
&= e^{\th|\dpr\L|}\, 
N \max_{0<\d<\r} \fra{e^{(\b\m_0(\r-\d)-\b f(\b,\r-\d)){|\Lambda|}} } 
{e^{(\b\m_0\r_0-\b
f(\b,\r_0))\,{|\Lambda|}}}\,e^{o(|\L|)}\eq(6.7)\cr}$$ 
where we have used the fact that the canonical thermodynamic limit
(with
Dirichlet boundary conditions) is uniform in the density in any closed
interval contained in $(0,\r_{cp})$, see [Ru69], and furthermore that
the maxima in \equ(6.7) are certainly achieved, uniformly in the
volume, for $k>\h |\L|$ for some $\h>0$.

In fact the maximum of the r.h.s. of (6.7) is achieved precisely at
$\d=0$ by
the convexity properties of the free energy because

$$\fra{e^{(\b\m_0(\r-\d)-\b f(\b,\r-\d))\,|\L|}}
{e^{(\b(\m_0\r_0-\b f(\b,\r_0))\,|\L|}}=
e^{-\b(f(\b,\r-\d)-f(\b,\r_0)-\dpr_\r f(\b,\r_0)
(\r-\d-\r_0))\,|\L|}\Eq(6.8)$$
Where we have used the relation between grand canonical pressure and
canonical free energy. We have assumed above that there are no phase
transitions in the sense that for each chemical potential considered
there is just one density possible. In particular this means that
$\r={\dpr P(\b,\m) \over \dpr \m}$ and $\m={\dpr f(\b,\r) \over \dpr
\r}$ and
therefore \equ(6.7), \equ(6.8) and \equ(6.5) imply \equ(6.4).
Existence of phase transitions affect the argument and the results in
the same way as they do in classical statistical mechanics and we do
not discuss the details.
\*

The above method is purely probabilistic and it relies on two
properties:
(1) the measure $d {\bf \Omega}$ is positive and (2) inequality
\equ(6.3). Therefore it applies also to the Bose gas case whenever
\equ(6.3) can be checked.  The inequality \equ(6.3) does not hold if
$z
e^{2\b B}>1$ so that new ideas seem necessary to treat the Bose
statistics case in spite of the important results of [Pa85] which, by
relying on Ginibre's representation of the Gibbs distributions,
provide a
proof of boundary conditions independence.
\*

\0{\bf 7. Beyond the small activity region for weakly interacting
fermions on a lattice.}
\numsec=7\numfor=1\*

In this section we discuss Fermionic lattice systems.  Using the
methods 
developed for the study of ground states in Fermionic systems 
allows one to deal with arbitrary $\mu_0$ and $\b$ at small coupling,
thus enlarging the region in which LD relations can be obtained, at
least if one assumes periodic boundary conditions, i.e.\ regarding $V$
as a torus.  (This is certainly {\it very restrictive} but we note
that
periodicity of boundary conditions is an assumption under which most
of
the existing results on fermions have been derived.)

Results can be obtained for all $\m_0$ (\ie also positive $\m_0$ which
is
not possible with the previous method).  {\it In the
continuum case, however, similar results can be derived only in
one-dimension or if an ultraviolet cut-off is imposed on $\vec k$}.
We
shall fix {\it a priori} $\m_0,\b$, with $\b>0$, and {\it then}
discuss the
large deviations for $v$ small enough (depending on the choice of
$\b,\m_0$).

One can wonder how is it that the low temperature techniques that,
even for lattice systems, have been proved useful only in
$1$-dimensional cases (meeting serious, so far insurmountable,
difficulties in higher dimension) can be of help here in an
arbitrary dimension.  However the low temperature techniques were
devised to treat the case $\b=+\io$, \ie ground states. We would
encounter the same difficulties in their application to large
deviations theory if we tried to employ them for the purpose of
analyzing LD at zero temperature.  Applying them at $\b^{-1}>0$ and at
weak interaction is, however, not difficult at least for lattice
systems.
\*

We consider a $d$-dimensional square box of even integer side $L$ with
periodic boundary conditions, and $\vec x=(n_1,...,n_d)$, with
$n_i=-L/2,\ldots, L/2$. The Hamiltonian is similar to \equ(1.1) with
the Laplace operator replaced by the discrete Laplace operator.  If
there is no interaction all properties of the system can be obtained
through the free {\it Schwinger function}, also called {\it
propagator},

$$g(\xx)={1\over\b |V|} \sum_{k_0,{\vec k}}{e^{i k_0
(x_0+0^-)+i{\vec k}{\vec x}}
\over -i k_0+E(\vec k)-\mu_0}\Eq(7.1)$$
where $x_0\in(-\b,\b)$ and $k_0={{2\pi\over\b}(m+{1\over 2})}$,
$m=0,\pm1,\ldots$, and $k_i={2\pi n_i\over L}$, $n_i=0,\pm1,\ldots,$
$\pm L/2$, $\vec k=(k_1,..,k_d)$, $L$ being the length of the side of
$V$ and $E(\vec k)=\sum_{i=1}^d(1-\cos k_i)$. The symbol $x_0+0^-$
means that the value is the limit from the left.

The ``propagator'', see for instance Sec. 2 in [BG90] or [BGPS92], is
defined in terms of the creation and annihilation operators for
fermions $a^\pm_\xx$ and of the free Hamiltonian
$T=\sum-(\fra12\D_{\xx_i}-\m_0)=\ig_V a^{+}_\xx(-\fra12
\D_\xx-\m_0)a^-_\xx\, d\xx$ as
\def\xx{{\V x}}
$$\eqalign{
g_+(\V\x,t)=&\hbox{Tr}\,e^{-(\b-t) T}a^-_{ \xx}e^{-t T}a^+_{ \xx'}
/\hbox{Tr}\,e^{-\b T}\cr
g_-(\V\x,t)=&\hbox{Tr}\,e^{-(\b-t) T}a^+_{
\xx}e^{-t T}a^-_{ \xx'} /\hbox{Tr}\,e^{-\b T}\cr}\Eq(7.2)$$
if $\V\x= \xx- \xx',\ t>0$, which are combined to form a single
function:
$$g({\V\x},t)=\cases{g_+({\V\x},t)&if $\b>t>0$\cr
-g_-(-{\V\x},-t)&if $-\b<t\le0$\cr}\Eq(7.3)$$
And \equ(7.1) is a well known Fourier representation of the function
in \equ(7.3), having replaced the temperature parameter $t$ with
$x_0$ for uniformity of notation. Note that the function $g(\V x,t)$
has to be regarded as a function of $t$ which is defined and periodic,
although not continuous,
in $[-\b,\b]$ (and of $\V x$ defined and periodic in $[-L,L]^d$): this
is important to keep in mind when studying Fourier transforms.
\*

{\it Note that $k_i\le 2\pi$ while $k_0$ is unbounded}: this
unboundedness
of the sum is an {\it ultraviolet problem} ``in the temperature
direction''
and restricting to a lattice has the advantage that no ultraviolet
problem
is present in the ``spatial direction''. The ultraviolet problem in
the
temperature direction is somewhat trivial: it is however essential to
have
no cut-off in the $k_0$ values.  Such cut--offs violate ``reflection
positivity'', \ie the resulting system is no longer Hamiltonian and
the
physical interpretation becomes unclear.
\*

We define the ``distance on the periodic box $V$'' as
$d_L(x_i)={L\over\pi}\sin ({\pi x_i\over L})$ and
$d_\b(x_0)={\b\over\pi}\sin ({\pi x_0\over \b})$ and
$\dd_{\b,L}(\xx)=(d_L(x_1),...,d_L(x_d), d_\b(x_0))$. This becomes the
ordinary distance in the limit $L\to\io$.
\*

{\it In the following we shall sometimes write $\sum_{\vec x}$ as $\ig
d\vec x$.}
\*

For all positive integers $\a$ there exists a
constant $C_\a$, independent of $\b,|V|,\mu_0$ such that
$$|g(\xx-\yy)|\le \nu^{-1}\,{C_\a\over 1+\nu^\a|
\dd_{\b,L}(\xx-\yy)|^\a}\Eq(7.4)$$
where $\nu^{-1}=1+1/\max\{-\mu_0,\b^{-1}\}$.
\*

In fact let $h(t)$ a $C^\io$ function which is $1$ for $t>2$
and $0$ for $t<1$; we can decompose the propagator $g$ into its
ultraviolet and infrared parts $g_u,g_i$; we write
$$\eqalign{
&g(\xx-\yy)\equiv g_u(\xx-\yy)+g_i(\xx-\yy)\defi\cr
&\defi {1\over\b |V|} \sum_{k_0,{\vec k}}h(k_0^2){e^{i k_0
(x_0+0^-)+i{\vec k}{\vec x}}
\over -i k_0+E(\vec k)-\mu_0}+{1\over\b |V|} \sum_{k_0,{\vec k}}
(1-h(k_0^2)){e^{i k_0
x_0+i{\vec k}{\vec x}}
\over -i k_0+E(\vec k)-\mu_0}
\cr}\Eq(7.5)$$
and we estimate them separately.

(1) First we check that the bound \equ(7.4) holds for
$g_i(\xx)$; if $n_0, n_1,...,n_d$ are positive integers then
$$|(d_L(x_1)^n_1...d_L(x_i)^{n_d}d_\b(x_0)^{n_0})\cdot g_i(\xx)|\le
{1\over\b |V|} \sum_{k_0,{\vec k}} |\partial^{n_1}_{k_1}...
\partial^{n_0}_{k_0} \hat g_i(\kk)|\Eq(7.6)$$
where $\partial_{k_i},\partial_{k_0}$ denotes the discrete derivative
on the lattice of the wave numbers $k$. Noting that the denominator in
\equ(7.1) is greater than $\max\{-\mu_0,\b^{-1}\}$, one obtains
for $g_i(\xx-\yy)$ the bound \equ(7.4) because the sum over $k_0$ is
finite and that over $\vec k$ tends to an integral over the Brillouin
zone which is also finite.

\0(2) The next relation that we need is that, for all $\a$ there is a
    constant $C_\a$ such that
$$|g_u(\xx-\yy)|\le
{C_\a\over 1+\n^\a \,|\dd_{\b,L}(\xx-\yy)|^\a}\Eq(7.7)$$
The ``only'' problem in checking \equ(7.7) is that the sum over $k_0$
is unbounded and the summand is $O({k_0^{-1}})$ so that the sum is
improperly convergent if $n_0=0$: indeed using
\equ(7.6) we see that if $|n_0|\ge 1$ the discrete derivative is
applied
either to the denominator (and gives a quantity bounded by $O(
k_0^{-2})$) or it is applied to $(1-h(k^2_0))$, and $\partial_{k_0}
h(k^2_0)$ has compact support.

Therefore we have to check that the lack of convergence only causes a
discontinuity of the function $g_u$ at the origin and $\pm\b$ and it does not
affect the large distance behavior. The non smoothness is just a
discontinuity hence it does not affect the size of $g_u$ at the
origin. A direct check that $g_u(\xx)$ is bounded is obtained by
rewriting $g_u$ as
$$\eqalign{
&\fra1{\b V}\sum_\kk h(k^2_0)\,\fra{e^{i k_0x_0+
\vec k\cdot\vec x} (E(\vec k)-\mu_0)}{(-ik_0+E(\vec k)-\m_0)(-ik_0)}+
\d_{\vec x,\vec 0}\,\fra{1}{\b}\,\sum_{k_0} (h(k^2_0)-1)\,\fra{e^{i
k_0x_0}}{-ik_0}+\cr
&+\fra12 \d_{\V x,\V0}-\ch(x_0)\d_{\vec x,\vec 0}
\defi \lis g(\vec x,x_0)-\ch(x_0) \d_{\vec x,\vec 0}\cr}
\Eq(7.8)$$
where $\ch(x_0)=1$ if $x_0\ge0$ and $\ch(x_0)=0$ otherwise;
furthermore $\d_{\vec x, \vec y}$ denotes the Kronecker delta on the
lattice points and we use that (recall that $g,g_u,g_i$ are periodic,
although not continuous,
functions of $x_0$ in $[-\b,\b]$).

$$\b^{-1}\sum_{k_0=(m+\fra12)2\p/\b} 
\fra{e^{ik_0 (x_0+0^-)}}{-i k_0}=-\fra12 {\rm
sign} (x_0)=\fra12-\ch(x_0)\ .\Eq(7.9)$$
The function $\lis g$ is defined by the r.h.s. of \equ(7.8); the first
two sums in \equ(7.8) are absolutely convergent (the second is a
finite
sum); hence $g_u$ is finite near $x_0=0$. Furthermore the Fourier
transform $\hat{{\lis g}}(\kk)$
of $\lis g$ is  such that its ``$L_1$--norm'' $(\b
V)^{-1}\sum_{k_0,\kk}
|\hat{{\lis g}}(\kk)|$ is uniformly bounded in $V$ because (if
$k_0=\p\b^{-1}n$ with $n$ integer)

$$\eqalign{
&|\hat{{\lis g}}(\kk)|\le \fra{2d+|\m_0|}{k_0^2}\ch(|k_0|>1)+
\fra{\ch(|k_0|<1)}{2\p\b ^{-1}}+ \fra\b2 \d_{k_0,0} \qquad{\tto}\cr
&(\b V)^{-1}\sum_{k_0,\kk}
|\hat{{\lis g}}(\kk)|\le c\,(1+|\m_0|)\b\cr}\Eq(7.10)$$
for a suitable constant $c$. Note that instead the Fourier transform
of $g$ {\it has a logarithmically divergent} $L_1$--norm in the above
sense because of the last term in \equ(7.8).
\*

After the above remarks on the nature of the infrared and ultraviolet
propagators we can go back to the problem of interest to us here.  We
study the Laplace transforms considered in 4, \equ(4.3), and write
$$\eqalign{
&{\rm Tr} e^{-\b(H_V-\mu_0 N)}e^{\b\l N_\L}=\cr
&=\sum_{n=0}^\io {1\over n!} (\b\l)^n\sum_{\vec y_1}.....\sum_{\vec
y_n}
\chi_\L(\vec y_1)...\chi_\L(\vec y_n)
{\rm Tr}[e^{-\b(H_V-\mu_0 N)}\r(\vec y_1)....\r(\vec
y_n)]\cr}\Eq(7.11)$$
where $\r(\vec x)=\psi^+_{\vec x}\psi^-_{\vec x}$.  Note that since
the fermions are on a lattice and in a finite container the series
over $n$ is a finite sum, by Pauli's principle.

It is well known that the partition function trace for a Fermionic
Hamiltonian can be written in terms of {\it Grassman's variables}
$\h^+_\xx\equiv\h^+_{{\vec x},x_0}$, $\h^-_\xx=\h^-_{{\vec x},x_0}$;
in particular, if $H_0$ is the kinetic energy (with the discrete
Laplacian)
$$\fra{{\rm Tr}[e^{-\b(H_V-\mu_0 N)}\r(\vec y_1)....\r(\vec y_n)]}
{{\rm
Tr}\,e^{-\b (H_V-\mu_0 N)} }=\fra{
\int P(d\h)e^{-U(\h)}\h^+_{{\vec y}_1,0}\h^-_{{\vec y}_1,0}...
\h^+_{{\vec y}_n,0}\h^-_{{\vec y}_n,0}}
{\int P(d\h)e^{-U(\h)}}
\Eq(7.12)$$
where $U(\h)=\e\int d\xx d\yy v({\vec x}-{\vec y})\d(x_0-y_0)
\h^+_{\xx}\h^-_{\xx}\h^+_{\yy}\h^-_{\yy}$,
$\int d\xx=\sum_{\vec x}\int_0^\b dx_0$. The ``integral'' $\int
P(d\h)\cdot$ is
defined on monomials of Grassman variables (and extended to general
functions by linearity) by the anticommutative Wick rule with
propagator
\equ(7.1).  Then we can write by \equ(7.11),\equ(7.12)
$$\media{e^{\l\b\,N_\L}}_{\m_0}=
%\lim_{M\to\io}
{\int P(d\h)e^{-U(\h)} e^{\b \l N_{\L}(\h)}
\over\int P(d\h)e^{-U(\h)}}\Eq(7.13)$$
where $N_{\L}(\h)=\sum_{\vec x} \chi_\L(\vec x)\h^+_{\vec x,0}
\h^-_{\vec x,0}$.
\*

\0{\it Remark:} had we considered instead of the ratio in \equ(4.3)
the quantity
$$\fra {{\rm Tr}\, e^{-\b(H_V-\mu_0 N+\l N_\L)}}{ {\rm Tr}\, e^{-\b
(H_V-\mu_0 N)}}=
{\int P(d\h)e^{-U(\h)} e^{\l\b\hat N_\L(\h)}\over
\int P(d\h)e^{-U(\h)}}\Eq(7.14)$$
where $\hat N_\L=\b^{-1}\int dx_0 \sum_{\vec x}\chi_\L(x)
\h^+_{\xx}\h^-_{\xx}$ we would have obtained a  different quantity
but estimates for it can be obtained in a way similar to the ones we
explain below: this is related to the discussion following \equ(2.4).
\*

In terms of Grassmanian integrals the logarithm of the Laplace
transform
of the probability \equ(4.1) divided by $\b |\L|$ is given by
$${1\over\b|\L|}\log \media{e^{\l\b\,N_\L}}_{\m_0}={1\over\b|\L|} \log
{
\int P(d\h)e^{-U(\h)} e^{\b\l N_\L}
\over \int P(d\h)e^{-U(\h)}}\Eq(7.15)$$
For a system of fermions on a cubic $d$-dimensional lattice with
Hamiltonian \equ(1.1), in which the Laplace operator is replaced by
the discrete Laplace operator, then we can proceed to developing a
{\it cluster expansion at small interaction} and proceed to studying
the large deviations probabilities following the same strategy of
Sec. 4, 5 above.
\*

In fact by the definition of {\it truncated expectation}

$$\log \int P(d\h)\, e^{\e A(\h)}\=\EE^T(e^{\e
A})\defi\sum_{n=0}^\io{\e^n\over n!}\EE^T(A;n)\Eq(7.16)$$
where $\EE^T(A;n)=\EE^T(A_1,A_2,\ldots, A_n)|_{A_i=A}$ with
$\EE(\cdot,\cdot,\ldots)$ is a suitably defined multilinear function
of its $n$ arguments, we can rewrite \equ(7.15) as

$${1\over \b|\L|}\EE^T(e^{-U}(e^{\l \b N_\L}-1))\Eq(7.17)$$
having exploited that Grassmanian variables anticommute (so that even
monomials in Grassmanian fields commute, {\it unlike the Fermionic
fields that generate them}). More explicitly we develop the
exponential in powers and obtain

$${1\over\b|\L|}[\sum_{n=0}^\io {1\over n!}\EE^T(-U+ \int
d\xx\,\chi_\L(\vec x)\,
\h^+_{{\vec x},0}\h^-_{{\vec x},0};n)-{1\over n!}\sum_{n=0}^\io
\EE^T(-U;n)]\Eq(7.18)$$
As usual one can represent the Grassmanian expressions
$$\eqalign{
&U(\h)=\int d\xx d\yy v({\vec x}-{\vec y})\d(x_0-y_0)
\h^+_{\xx}\h^-_{\xx}\h^+_{\yy}\h^-_{\yy}\cr
&N_\L(\h)=\b\,\l\,\ig
\chi_\L(\vec y)\d(y^0) \h^+_\yy \h^-_\yy\,d\yy\cr}\Eq(7.19)$$
by the diagrams or ``{\it graph elements}'' in Fig.1.
\*

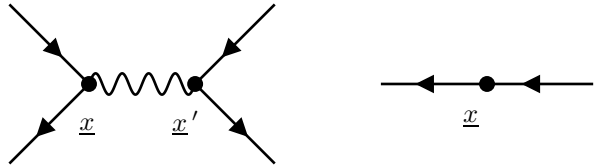
\figini{grafici}
\8</origine1assexper2pilacon|P_2-P_1| { >
\8<4 2 roll 2 copy translate exch 4 1 >
\8<roll sub 3 1 roll exch sub 2 copy atan >
\8<rotate 2 copy exch 4 1 roll mul 3 >
\8<1 roll mul add sqrt } def >
\8<>
\8</punta0 { % h punta0 >
\8<0 0 moveto dup dup 0 exch 2 div lineto 0 >
\8<lineto 0 exch 2 div neg lineto >
\8<0 0 lineto fill stroke } def >
\8<>
\8</freccia0 { % l freccia0 >
\8<0 0 moveto 0 2 copy lineto stroke exch 2 >
\8< div exch translate 7 punta0 >
\8<stroke } def >
\8<>
\8</freccia{ % x1 y1 x2 y2 freccia >
\8<gsave origine1assexper2pilacon|P_2-P_1| >
\8<freccia0 grestore} def >
\8<>
\8</onda0 { >
\8<dup dup dup 10 div round div dup 100 div >
\8<exch 1 exch div 1 mul 360 mul >
\8<3 1 roll exch 0 3 1 roll 0 0 moveto >
\8<{dup 2 index mul sin 4 mul lineto} >
\8<for stroke pop pop } def >
\8<>
\8</onda { % x1 y1 x2 y2 onda>
\8<gsave origine1assexper2pilacon|P_2-P_1| >
\8<onda0 grestore} def >
\8<>
\8</punto { % x y punto>
\8<gsave 3 0 360 newpath arc fill stroke grestore} def>
\8<>
\8<gsave>
\8</P1 {10 10} def>
\8</P2 {40 40} def>
\8</P3 {10 70} def>
\8</P4 {110 10} def>
\8</P5 {80 40} def>
\8</P6 {110 70} def>
\8</P7 {190 40} def>
\8</P8 {150 40} def>
\8</P9 {230 40} def>
\8<>
\8<P3 P2 freccia>
\8<P2 P1 freccia>
\8<P2 P5 onda>
\8<P6 P5 freccia>
\8<P5 P4 freccia>
\8<P9 P7 freccia>
\8<P7 P8 freccia>
\8<>
\8<P2 punto P5 punto P7 punto>
\8<grestore>
\figfin

\eqfig{230pt}{70pt}
{\ins{35pt}{27pt}{$\xx$}
\ins{70pt}{30pt}{$\xx'$}
\ins{180pt}{30pt}{$\xx$}
}{grafici}{}
\*
\0{Fig.1: \nota The lines ``entering (exiting) a point'' $\xx$
represent the
Grassmanian fields $\ps^+_\xx$ (respectively $\ps^-_\xx$); the wavy
line connecting the dots marked $\xx,\xx'$ represents
$v(\vec x-\vec x')\d(x^0-x^{0\prime})$ while the dot in the second
diagram represents $\ch_\L(\vec x)\d(x_0)$.}
\*

We can call the pair $P=(\xx,\xx')$ or the single point $Q=\xx$
in Fig.1 a {\it cluster}. If
$$\eqalign{
&dP=d\xx d\xx',\ dQ=d\xx,\ V(P)=v(\vec x-\vec
x')\d(x^0-x^{\prime0}),\cr
&\ch_\L(Q)=\ch_\L(\vec x)\d(x^0),\
\h(P)=\h^+_\xx\h^{-_\xx}\h^+_{\xx'}\h^-_{\xx'},\
\h(Q)=\h^+_\xx\h^{-_\xx}\cr}
\Eq(7.20)$$
and if we set, as above, $\ig d\xx\cdot\defi\ig_0^\b dx^0\,d
\vec x\cdot \,\defi\,
\ig_0^\b dx^0\sum_{\vec x}\,\cdot$ the \equ(7.18) takes the form
$$\eqalign{
&\sum_{n_1,n_2\atop n_2\ge 1}\ig (\prod_{i=1}^{n_!}
V(P_i)\,dP_i)\,\cdot\,(\prod_{j=1}^{n_2}
\ch_\L(Q_j) \,dQ_j)\cdot \EE^T\big((\prod_{i=1}^{n_!}
\h(P_i))\,(\prod_{j=1}^{n_2}\h(Q_j));n_1+n_2\big)\cr}\Eq(7.21)$$
simply by the multilinearity of the truncated expectations, see
[Le87],
[BGPS94]. The sum over $n_2$ starts from $n_2=1$ because the terms
with $n_2=0$ cancel because for $\l=0$ \equ(7.17) vanishes.

We now imagine to join or ``{\it contract}'' (``Wick's contraction'')
pairs of lines associated with clusters (defined above) and with
matching orientations in such a way that the collection of lines thus
obtained, including the wavy lines, form a {\it tree graph} connecting
all points of the clusters. The geometric object so built is called a
``spanning tree'' and contains $2(n_1+n_2-1)$ solid lines and $n_1$
wavy lines. In the graph there will be, therefore, $2m\defi 4
n_1+2n_2-2(n_1+n_2-1)=2(n_1+1)$ lines left ``uncontracted''.
\*

A remarkable algebraic expression for  the truncated Fermionic
expectations $\EE^T$ in \equ(7.21) can be developed (``Lesniewsky's
expansion'') as
$$\sum_{T= \,spanning\ tree}\ig
dr_{T}(\ttu)\,(\prod_{l\in T}
g(\x_l))\cdot(\det G^{T}(\ttu))\Eq(7.22)$$
where:

\0(i) $\x_l=\xx-\yy$ if the line $l\in T$ joins $\xx$ and $\yy$,\\
(ii) $\ttu$ is a family of $2m$ ``interpolation parameters in
$[0,1]$,\\
(iii) $r_T(\ttu)d\ttu$ is a probability measure on the interpolation
parameters whose structure is inessential but for the fact that $\ig
r_T(\ttu)d\ttu=1$,\\
(iv) $G^T$ is a $m\times m$ matrix obtained by
considering the $m$ points $\tilde \xx_1,\ldots,\tilde \xx_m$ or
$\tilde \yy_1,\ldots,\tilde \yy_m$ into which enter, respectively
exit, uncontracted lines and setting $G^T_{ij}= t_{ij} \cdot g(\tilde
\xx_i-\tilde \yy_j)$ with $t_{ij}$ functions of the interpolation
parameters such that $t_{i,i'}=\uu_i\cdot\uu_{i'}$ for a suitably
defined unit vectors $\uu_i\in R^m$. The latter property of the unit
vectors is the only property that concerns us here.
\*

The above formula is very convenient because of the Gramm--Hadamard
inequality\annota{5}{This inequality concerns matrices having the form
$M_{i,j}=(u_i\cdot v_j)_H$ where $u_i,v_j$ are unit vectors in a
Hilbert space $H$ and $(u\cdot v)_H$ denotes the scalar product in $H$
and it states that the determinant of $M$ is bounded by $1$ in
absolute value. It has, essentially, the simple geometric meaning that
the volume of a parallelepiped with sides of length $1$ cannot exceed
$1$. An immediate consequence, that we employ here, is that matrices
$M_{i,j}=(u_i\cdot v_j)_H\,(U_i\cdot V_j)_K$ with $u_i,v_j$ unit
vectors in a Hilbert space $H$ and $U_i,V_j$ unit vectors in a Hilbert
space $K$ also have determinant bounded by $1$ (one just applies the
previous inequality in the Hilbert space $H\times K$).} which applies
precisely to matrices that have the form of $G^T$ and bounds their
determinant (for all $\ttu$) in terms of the $m$--th power of the
$L_1$--norm of the integrand in \equ(7.1):

$$|\det G^T(\ttu)|\le c_1^m \Eq(7.23)$$
{\it provided the norm is finite}. However, as remarked in connection
with \equ(7.9), \equ(7.10) there is no cut--off on the momentum $k_0$
and the norm diverges (``logarithmically'').  Here we hit a problem
which is serious because apparently the best estimate that we can find
for this determinant using just the boundedness of the matrix elements
is like \equ(7.23) but with an extra $m!$, which would be a
disaster.

Suppose, {\it temporarily}, that the propagator $g(\xx-\yy)$
is replaced by $g_i(\xx-\yy)$ (\ie we impose an
{\it ultraviolet cut-off} to the model) so that 
the inequality
\equ(7.23) holds: then we can quickly develop a proof of a LD result.
In fact, assuming \equ(7.23) valid with some constant $c_1$, we can
bound the $n_1,n_2$ term of the series \equ(7.21) by

$$\fra{|\l|^{n_2}}{n_1!n_2!} (\ig {d\xx \over \nu}{C_\a\over 1+\nu^\a|
\dd_{\b,L}(\xx)|^\a})^{n_1+n_2-1}\cdot(\sum_{\vec x}
v(\vec x))^{n_1}\cdot c_1^{n_1+1}
\n^{-1}\ch_\L(\vec y_1)\d(y_1)\Eq(7.24)$$
if we do not perform the integral over the center $\yy_1$ of the
cluster $Q_1$ which certainly exists because $n_2\ge1$: this expresses
the cancellation in \equ(7.18) between the terms arising from the
first truncated expectation when one considers only the contributions
from $-U$ out of the $2^{n}$ that are generated by expanding the sums
in
the first expectation (\ie the terms corresponding to $n_2=0$).
\*

We see that the difference \equ(7.18) can be given the form

$$\sum_{n1,n2\ge0\atop n_2\ge1} \l^{n_2}\ig W(X,Y)
\ch_\L(Y)\,dX\,dY\Eq(7.25)$$
where $W(X,Y)$ is a translation invariant function expressible (as
shown implicitly above) as a sum of a large number of tree graphs
connecting the points in $Y\cup X$ with $X=(\xx_1,\ldots, \xx_{n_1}),
Y=(\yy_1,\ldots, \yy_{n_2})$. Furthermore if one fixes the point
$\yy_1\in Y$, say, the integral over the remaining points
$X,Y^{(1)}=Y/\yy_1$ is bounded, for a suitable choice of $c_2$, by

$$\ig |W(X,\yy_1,Y^{(1)})|\,dX\,dY^{(1)}\le \e^{n_1-1}
c_2^{n_1+n_2}\Eq(7.26)$$
having used Cayley's tree--counting formula to take into account of
the
sum over the spanning trees and having set $\e=\sum_{\vec x}|v(\vec
x)|$.
We shall show in Appendix A1 that \equ(7.26) holds as well without
imposing an ultraviolet cut--off.

Hence we can write the difference in \equ(7.18) with the notations of
4, see \equ(4.7), as

$$\b (P_\ell(\b,\m_0;\l)-P_\ell(\b,\m_0)=
\ell^{-d}\sum_{n1,n_2\ge0\atop n_2\ge1} \l^{n_2}
\ig W(X,Y)\,\ch_\L(Y) dXdY\Eq(7.27)$$
which has the same structural properties as the first of \equ(4.7) and
we can therefore proceed in the same way as in 4, 5, with
\equ(7.26) playing the role of \equ(3.3), to derive
for a system of fermions on a lattice:
\*

\0{\it (V) Large deviation property 5:}
\*
\0{\it Let $\b,\m_0$ be fixed arbitrarily
and let $\r_0$ be the corresponding density (so that
$\r_0=\r(\beta,\mu_0)$).  Suppose that the $\sum_{\vec x} |v(\vec
x)|=\e$ is small enough, depending on $\b,\m_0$. If the side $\ell$ of
the box $\L$ tends to infinity and, correspondingly, the container
side also tends to infinity so that $L/\ell\to\io$ then

$$\lim_{\ell\to\io} {\ell^{-d}}\log\sum_{\r\in[a,b]} \Pi(\r\ell^d)=
\max_{\r\in{[a,b]}} \, -\b\,\D F(\b,\r,\r_0)\Eq(7.28)$$
This holds if the interval $[a,b]$ is
contained in an interval $[-\d_0(\b,\m_0),$ $\d_0(\b,\m_0)]$ centered
at
$\r_0$ with $\d_0(\b,\m_0)=c_2^{-1}\n^{(d+2)}>0$ small enough.}

\*
We note that we can replace $v(\xx)$ by $|v(\xx)|$ in the bounds
above:
hence repulsivity of the potential is not necessary and has
not been mentioned in the statement above. This reflects an essential
difference between lattice and continuum systems, see concluding
remark
(7).
\*

To conclude the proof we refer to Appendix A1 where we prove
\equ(7.25). The inequality \equ(7.26) is studied in [BGPS94]: sec. 3
of that reference is in fact dedicated to a detailed analysis of the
ultraviolet problem in a case which is more involved than the present
one. See appendix A1 below.

\*
\0{\bf 8. Conclusions.}
\numsec=8\numfor=1\*

(1) In the cases of Boltzmann statistics with arbitrary $\b,\m_0$ or
    in the case of Bose statistics with $z e^{2\b B}<1$ the
    Ginibre representation allows us to regard the quantum gas as a
    classical gas of contours, at least for the purposes of computing
    the partition function or the probability distribution of the
    number of particles in a given region. This is so  because
    the integration measure over the trajectories is with a positive
definite measure $d{\bf \O}$.
Hence we can apply the arguments used in the general derivation of
    the large deviations formulae from the theory of the equivalence
    of ensembles of [La72] as done in the classical case in
    [Ol88]. The applicability of such methods found an early
    application in the work of Ginibre, see [Gi72] p. 362. Naturally
    we called the results ``weaker'' because we did not get the
    ``total'' control on the free energy and the accurate treatment of
    the finite size effects that \equ(3.3) and \equ(3.6) allow us to
    derive, but the extra generality is nevertheless remarkable.

(2) The above analysis admits a straightforward extension to
    electron--phonon systems because of the remarks in [GGV70], see
    also [Gi71], p. 420.

(3) We have also implicitly obtained that the asymptotic behavior of
    \equ(4.3) (when $\ell\to\io$, $L\to\io$ and $\L/\ell \to\io$) is
    the
    {\it same to leading order} as that of the r.h.s. of \equ(4.1), in
    the analyticity regions considered.

(4) The results for Fermi systems, aside from  being restricted to
    lattice systems, considered in sec. 7 cover
    in the fernionic case a rather different region of the
    $(\b,\m_0)$--plane, compared to the ones derived in sec. 4. They
    are
valid
 at any fixed $\b,\r$ (or at any
    $\b,\m_0$) provided the coupling is small enough. Thus sec. 7
    provides a stronger connection with the results for free fermions
    in [LLS99] which are implied by our results in the interval of
    density $(-\d(\b,\m_0),\d(\b,\m_0))$ in which the latter hold.

(5) It may seem that the obstacle to deriving a LD relation in the
    case of the Bose or Fermi statistics and in the region where
    genuinely quantum phase transitions can occur is related to the
    problem of showing boundary condition independence of the main
    extensive thermodynamic quantities. However this problem has been
    solved in many cases: in some cases, indeed, the technical
    estimates of its solution imply, see [Ro70] and [Gi71] p. 365, our
    equations \equ(3.4) and \equ(6.3), which have been the basis of
    our results in the first six sections of this paper; but in other
    cases, and we think of the paper [Pa85], boundary conditions
    independence can be proved for very general interactions (\eg
    bosons with just superstable, if $d<4$, or superstable and
    repulsive
    interactions, if $d\ge4$): but the result does not imply \equ(6.3)
    and therefore it seems that LD needs more than just boundary
    condition independence. Perhaps \equ(6.3) has to be replaced by
    estimates that involve all particles coherently, as in [Pa85],
    rather than single ones as it is essentially the case in the
    proofs of \equ(6.3) (which depend on the finite size of the
    individual closed brownian paths).

(6) The results in sec. 7 are valid for small interactions: how small
    depends on $\b$ and $\mu_0$. The only case in which a similar
    result
    can be found for a continuum system is the 
    $d=1$ case. If we consider a $1$--dimensional continuum system we
    can proceed in the same way after replacing sums over $\vec x$
    with integrals and using that the kinetic energy grows
    quadratically at $\io$ so that the integral replacing the sum in
    \equ(7.4) is still convergent: this is peculiar to $d=1$ so that
    the result does not extend to higher dimension {\it unlike the
    lattice case}. Of course if an ultraviolet cut--off is imposed on
    $\vec k$ the above analysis carries over to higher dimension:
    since ultraviolet cut--offs on the spatial momenta $\vec k$ {\it
    preserve reflection positivity} the result has some interest
    because it expresses properties of systems which are still
    Hamiltonian in spite of a ``strange'' form of the kinetic energy
    for large $\vec k^2$.

(7) The key bound \equ(7.26) has been rederived here (see Appendix)
    but it is implicit in the analysis in [BGPS94], Sec. 3, where the
    more complex problem of the ultraviolet stability in one
    dimensional Fermionic systems has been treated in detail. We have
    not simply referred to [BGPS94] for the sake of completeness (the
    result is somewhat hidden there as the concern was about different
    topics). It would be interesting to prove that the inequality
    \equ(7.23) holds in spite of the fact that the $L_1$ norm of the
    Fourier transform of $g$ is infinite; or, alternatively, to show
    that it does not hold so that the proof that we reproduce in
    Appendix A1 for \equ(7.25), \equ(7.26), {\it independent of} the
    validity of the bound \equ(7.23), is in a sense optimal.

\*\*

\0{\bf Acknowledgements:} We are indebted to H.T. Yau and H. Spohn for
several discussions and examinations of alternative methods.
\*

\0{\bf  A1. Appendix: Proof of \equ(7.25), \equ(7.26).}
\numsec=1\numfor=1
\*

We shall represent the Grassmanian fields $\h$ in \equ(7.13) as
$\lis\h^\pm+\ps^\pm$ where $\lis\h$ has propagator $\lis g(\xx-\yy)$
as defined in \equ(7.8) while $\ps$ has propagator, see \equ(7.8),
$-\ch(x_0)\d_{\vec x,\vec0}$.

The Grassmanian integrals in \equ(7.13) will be written as double
integrals: for instance the numerator in \equ(7.13) will be written

$$\ig P(d\lis\h)\ig P(d\ps) e^{-U(\lis\h+\ps)+\b\l
N_\L(\lis\h+\ps)}\Eqa(A1.1)$$
The difference with the corresponding treatment in [BGPS94] is that we
use the simplicity of the propagator of $\ps$ to perform explicitly
the integration over $\ps$. The integration over $\ps$ in \equ(A1.1)
can be performed via the formula
\equ(7.16) and we replace \equ(7.17) by

$$\fra1{\b|\L|} \Big[
\EE_{\lis \h}^T \Big(\EE^T_\ps(e^{-U(\lis\h+\ps)+\b\l
N_\L(\lis\h+\ps)})\Big) -
\EE_{\lis \h}^T
\Big(\EE^T_\ps(e^{-U(\lis\h+\ps)})\Big)\big]\Eqa(A1.2)$$
The evaluation of $\EE^T_\ps(e^{-U(\lis\h+\ps)+\b\l
N_\L(\lis\h+\ps)})$ is similar to that of $\EE^T
(e^{-U(\h)+\b\l N_\L(\h)})$ in \equ(7.18) through \equ(7.19) above.
The $\EE^T_\ps$ will be a function expressed as a sum of
Grassmanian monomials in the fields $\lis\h$: each such monomial will
arise as a ``value'' of a suitable graph according to a (classical)
procedure that we proceed to describe.
\*

Since $U$ and $N_\L$ depend upon $\lis\h+\ps$ we develop the monomials
in $U$ and $N_\L$ obtaining a sum of $20$ monomials
in $\ps$ with coefficients that depend on $\lis\h$ ($16$ from $U$ and
$4$ from $N_\L$): each of such terms can be represented graphically by
a graph like those in Fig. 1 in 7 in which one or more of the solid
lines is replaced by a dashed line (hence we get $16$ different graphs
from the first and $4$ different ones from the second):
the dashed lines represent the Grassmanian fields $\ps$ while the
solid lines represent the fields $\lis\h$.

For instance a graph with $3$ solid lines and one dashed line will
represent $\lis\h^+_x\lis\h^-_x\lis\h^+_{x'}\ps_{x'}^-$ if in Fig. 1
above a dashed line has replaced the solid line entering $x'$.
\*

Keeping the notation introduced in 7 we can say that while in the
approach attempted in \S7 we had just two types of clusters (denoted
$P$ or $Q$) we now have $20$ different types of {\it clusters}
symbolically represented by ``{\it graphs elements}'' like those in
Fig. 1 but with one or more solid lines replaced by dashed ones. The
different types of graphs elements will be labeled with a label
$j=1,2,\ldots,16, 17,\ldots,20$: the last $4$ will be the ones
generated by the second graph element in Fig. 1 (\ie by the graph
element with two lines only).
\*

The truncated expectation $\EE^T_\ps$ is computed according to the
usual rules: namely we must form all possible graphs $\g$ with
$n=\sum_{j=1}^{20} n_j$ clusters, of which $n_j$ are of type $j$, in
which the dashed lines are pairwise contracted or ``paired'' to form
oriented dashed lines connecting nodes of graph elements {\it so that
no dashed line is left out ``unpaired'' and the graph obtained is a
connected graph $\g$}. The wavy lines present in the first $16$ graph
elements must taken into account in checking whether or not a graph is
connected.

The ``value'' of a graph will be a monomial in the Grassmanian
variables $\lis\h$ rather than just a number as it was the case in
7.
\*

Consider a graph $\g$ formed from $n_j$ clusters $P_{j,i}$,
$i=1,\ldots,n_j$, of type $j=1,\ldots,16$ or $Q_{j,i}$,
$i=1,\ldots,nj$ of type $j=17,\ldots$, with a total of $2n_{dash}$ of
dashed lines. Then $\g$ will represent a monomial in the fields
$\lis\h$ that we denote $\p_\g(PQ)$ and which is the product of the
fields $\lis\h$ associated with the solid lines (all of which remain
``unpaired'' because pairing only concerns the dashed lines, \ie the
fields $\ps$) of the clusters.

Since the propagator of a pair of fields $\ps^-_x,\ps^+_{x'}$ is
$-\d_{\vec x-\vec x'}\ch(x_0-x'_0)$ a factor $(-1)^{n_{dash}}$ has to
be added {\it together with all the Kronecker deltas and all the
$\ch(\cdot -\cdot)$ functions expressing that the various $x_0-x_0'$
differences must have a definite sign}.

We call $\D(\g)$ the value of the product of such quantities: it has
value $0,\pm1$ and we denote $P_j,Q_j$ the collection of the nodes of
the clusters $P_{j,i}$ or $Q_{j,i}$ respectively. In conclusion the
value of the graph $\g$ can be written as

$$\ig \Big(\prod_{j=1}^{16} (\prod_{i=1}^{n_j} V(P_{j,i}))
\fra{dP_{j}}{n_j!} \Big)
\cdot \Big(\prod_{j=17}^{20} (\prod_{i=1}^{n_j} \l \ch_\L(Q_{j,i}))
\fra{dQ_{j}}{n_j!}\Big)\p_\g(PQ) \,\D(\g)\Eqa(A1.3)$$
\*

{\it The key remark is that the graph $\g$ cannot contain any closed
loop other than the ones formed by contracting lines that exit from
the same graph element (giving rise to what are usually called
``tadpoles'') because the $\ch$ functions in the propagator of $\ps$
force the $x_0$'s of the various nodes of the $20$ graph types to be
in increasing order in the direction of the arrows drawn on the dashed
lines that are contracted.}
\*

Note in fact that only the graph elements corresponding to the first
$16$ types contain fields computed at exactly equal times (a necessary
condition for the possibility of loops).

Therefore the number of graphs $\g$ which are all trees is not bigger
than $(\sum_{j=1}^{20} n_j)!$ and we can use $(\sum_{j=1}^{20}
n_j)!/\prod_j n_j!< 20^{\sum_j n_j}$ in estimating the number of
different monomials that we can get in the construction. The number of
graphs which are not trees because of tadpoles is not much bigger
(being bounded by $6^n n!$).
\*

In this way we express $\EE^T_\ps(e^{-U(\lis\h+\ps)+\b\l
N_\L(\lis\h+\ps)})=\log \EE_\ps(e^{-U(\lis\h+\ps)+\b\l
N_\L(\lis\h+\ps)})$ as a sum of monomials in the fields
that we can ``{\it clusters}'' and which can be represented
graphically in a way similar to the one used in Fig.1, Sec. 7, to
represent the monomials in $-U(\h)+\b \l N_\L(\h)$; namely as in the
following figure Fig.2 \ifnum\mgnf=1 and analytically as\fi

\figini{gr2}
\8</origine1assexper2pilacon|P_2-P_1| { >
\8<4 2 roll 2 copy translate exch 4 1 >
\8<roll sub 3 1 roll exch sub 2 copy atan >
\8<rotate 2 copy exch 4 1 roll mul 3 >
\8<1 roll mul add sqrt } def >
\8<>
\8</punta0 { % h punta0 >
\8<0 0 moveto dup dup 0 exch 2 div lineto 0 >
\8<lineto 0 exch 2 div neg lineto >
\8<0 0 lineto fill stroke } def >
\8<>
\8</freccia0 { % l freccia0 >
\8<0 0 moveto 0 2 copy lineto stroke exch 2 >
\8< div exch translate 7 punta0 >
\8<stroke } def >
\8<>
\8</freccia{ % x1 y1 x2 y2 freccia >
\8<gsave origine1assexper2pilacon|P_2-P_1| >
\8<freccia0 grestore} def >
\8<>
\8</onda0 { >
\8<dup dup dup 10 div round div dup 100 div >
\8<exch 1 exch div 1 mul 360 mul >
\8<3 1 roll exch 0 3 1 roll 0 0 moveto >
\8<{dup 2 index mul sin 4 mul lineto} >
\8<for stroke pop pop } def >
\8<>
\8</onda { % x1 y1 x2 y2 onda>
\8<gsave origine1assexper2pilacon|P_2-P_1| >
\8<onda0 grestore} def >
\8<>
\8</punto { % x y punto>
\8<gsave 3 0 360 newpath arc fill stroke grestore} def>
\8<>
\8<gsave>
\8</X {90} def>
\8</R {30 } def>
\8</A {60} def>
\8</P1 {   A     cos      R mul X add        A     sin R mul X add   }
def>
\8</P2 { 2 A mul cos      R mul X add      2 A mul sin R mul X add   }
def>
\8</P3 { 3 A mul cos      R mul X add      3 A mul sin R mul X add  }
def>
\8</P4 { 4 A mul cos      R mul X add      4 A mul sin R mul X add  }
def>
\8</P5 { 5 A mul cos      R mul X add      5 A mul sin R mul X add  }
def>
\8</P6 { 6 A mul cos      R mul X add      6 A mul sin R mul X add  }
def>
\8<>
\8</PP1 {   A     cos 3 R mul mul X add   A     sin 3 R mul mul X add}
def>
\8</PP2 { 2 A mul cos 3 R mul mul X add 2 A mul sin 3 R mul mul X add}
def>
\8</PP3 { 3 A mul cos 3 R mul mul X add 3 A mul sin 3 R mul mul X add}
def>
\8</PP4 { 4 A mul cos 3 R mul mul X add 4 A mul sin 3 R mul mul X add}
def>
\8</PP5 { 5 A mul cos 3 R mul mul X add 5 A mul sin 3 R mul mul X add}
def>
\8</PP6 { 6 A mul cos 3 R mul mul X add 6 A mul sin 3 R mul mul X add}
def>
\8<>
\8<X X R 20 add 0 360 arc stroke>
\8<>
\8<P1 punto P2 punto P3 punto P4 punto>
\8<P5 punto P6 punto>
\8<>
\8<P1 PP1 freccia>
\8<PP2 P2 freccia>
\8<P3 PP3 freccia>
\8<PP4 P4 freccia>
\8<P5 PP5 freccia>
\8<PP6 P6 freccia>
\8<grestore>
\figfin

\midinsert
\eqfig{175pt}{175pt}
{\ins{77pt}{97pt}{$\st\V z_1\ldots \V z_p$}
\ins{106pt}{90pt}{$\st \xx_1$}
\ins{100pt}{107pt}{$\st \xx_2$}
\ins{75pt}{107pt}{$\st \xx_3$}
\ins{65pt}{90pt}{$\st \xx_4$}
\ins{75pt}{75pt}{$\st \xx_5$}
\ins{102pt}{75pt}{$\st \xx_6$}}
{gr2}{}

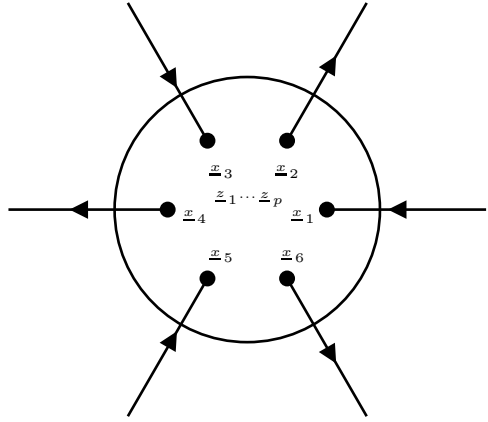
\0Fig.2: {\nota Illustration of a contribution to $\EE^T_\ps$ in
\equ(A1.2) from a cluster generated by collecting all graphs $\g$
which have $6$ external lines and contain $p$ graph elements of type
$>16$ with nodes at $\V z_1,\ldots,\V z_p$. These are graphs that
contribute with a monomial of the form $\lis\h_{\xx_1}^+\lis
\h_{\xx_3}^+\lis\h_{\xx_5}^+\lis \h_{\xx_2}^-\lis \h_{\xx_4}^-\lis
\h_{\xx_6}^-$.}
\*
\endinsert
\ifnum\mgnf=0 and analytically as\fi
$$\eqalign{
&\l^p \ig V(\xx_1,\ldots,\xx_n,\xx'_1,\ldots,\xx'_n,\V z_1,\ldots,
\V z_p)\,(\prod_{i=1}^p \ch_\L(\V z_i))\cdot\cr
&\cdot \lis\h_{\xx_1}\ldots\lis\h_{\xx_n}\ldots
\lis\h_{\xx'_1}\ldots\lis\h_{\xx'_n} (\prod_{i=1}^n d \xx_i)
(\prod_{i=1}^p d \V z_i)\cr}\Eqa(A1.4)$$
The $V$ function is obtained by collecting and summing contributions
from graphs $\g$ which lead to a monomial of the degree $2n$ in the
$\lis\h$ fields and contain $p$ graph elements with only two
lines. Such contributions will admit, if $p\ge 1$, the bound$$\ig
|V(\xx_1,\ldots,\xx_n,\xx'_1,\ldots,\xx'_n,\V z_1,\ldots,
\V z_p)|\,d\xx d\xx' \,d\V z_2\ldots d\V z_p< c_2^{n+p}\Eqa(A1.5)$$
Note that the kernel $V$ may contain some delta functions (or
Kronecker deltas) as it happens already in the corresponding
representation for $-U$ and $\l\b N_\L$ described after \equ(7.12).

We do not discuss further details because the technique has been used
many times in the literature and in particular in [BGPS94]. The
validity of the bound in \equ(A1.5) is an immediate consequence of the
fact that all the graphs that contribute to $V$ are connected. This
means that in the graph value one can associate a factor $v(\vec
x-\vec x')\d_{x_0-x_0'}$ to each wavy line that connects $\xx$ and
$\xx'$ or a factor $\d_{\vec x,\vec x'}$ (while no decay is present in
the $x_0-x_0'$) if the points are linked by a dashed line. {\it The
lack of decay in the $x_0$ variables is not a problem because the
$x_0$'s vary in a finite interval $[0,\b]$.} The combinatorics
requires some care and the analysis is based on the remark following
\equ(A1.4) and is also discussed in [BGPS94], for instance.
\*

We can now proceed to the integrations over the $\lis\h$ fields.

The procedure is the same as the one above. A minor difference is
that we have now {\it infinitely many graph elements}: namely all the
ones of the form illustrated in Fig.1 and in
\equ(A1.4).  This is not really a big difference because we have the
bound \equ(A1.5) which will allow us to control the proliferation of
graph elements.

{\it This time we can get, however, graphs with loops} and we cannot
control their number unless we are willing to obtain bounds that grow
factorially with the size of the graph. To avoid the appearance of
factorials we must collect many graphs together; or we must compute
the
expectations by using Lesniewsky's formula to avoid getting
combinatorially large contribution.
\*

{\it The idea is that this time the propagator $\lis g$ of the fields
$\lis \h$, that we have to integrate, has a finite $L_1$ norm, as
pointed out in the comment to \equ(7.10), so that we can apply the
formula \equ(7.23).}
\*

Furthermore the second truncated expectation appearing in \equ(A1.2)
will not contain any term with $p>0$ (of course), but it will
otherwise coincide with the expression of the first expectation. So
that the difference of the two will be given simply by the sum of the
terms with $p>0$. And combining \equ(A1.4) with \equ(7.23) one gets an
expression like \equ(7.25) and the bound \equ(7.26) with the symbols
$\xx_1,\ldots$ and $\V z_1,\ldots$ taking the place of the $X$ and $Y$
respectively.

Logically we avoided proving the estimate \equ(7.23) for the full $\h$
integration, and we still do not know whether such an estimate is
really valid. This has been done by performing {\it exactly} the
(relatively) simple integral over the component $\ps$ of the
Grassmanian field $\h=\lis\h+\ps$ which is responsible for the
summability problem described in the comments to
\equ(7.23). In this way it becomes clear that the logarithmic
divergence of the $L_1$ norm of the Fourier transform of the
propagator cannot really cause problems in the estimates that we need
for the purposes of proving \equ(7.26) and the LD relation 5 of 7.
\*

It is worth stressing that the problem that we studied in this
appendix (originally solved in [BGPS94]) is usually considered
``irrelevant'' in most of the literature: in condensed matter Physics
it is very common to find that the infrared problems are directly
attacked by {\it assuming that the propagator contains an ultraviolet
cut--off} both on the $\vec k$ and the $k_0$ components.\annota{5}{In the
present context this means taking as propagator the $g_i$ in \equ(7.5)
which of course allows to derive the \equ(7.23) with no effort and
save the analysis of this appendix.} This is done even in works in
which the authors are aware that a cut--off on $k_0$ is very difficult
to interpret because it breaks the reflection positivity of the
theory, \ie makes it ``non Hamiltonian''.
\*
\ifnum\mgnf=1\vfill\eject\fi

\0{\bf References.}
\*

\0[BG90] {\sl Benfatto, G., Gallavotti, G.}: {\it Perturbation
theory of the Fermi surface in a quantum liquid. A general quasi
particle formalism and one-dimensional systems}, Journal of
Statistical Physics, {\bf 59}, 541--664, 1990.

\0[BGPS94] {\sl Benfatto, G., Gallavotti, G., Procacci, A., Scoppola,
B.:}
{\it Beta function and Schwinger functions for a many body system in
  one dimension. Anomaly of the Fermi surface.}, Communications in
  Mathematical Physics, {\bf 160}, 93--172, 1994.

\0[GGV70] {\sl Gallavotti, G., Ginibre, J., Velo, G.}:
{\it Statistical mechanics of the electron--phonon system}, Lettere al
Nuovo Cimento, {\bf4}, 1293--1997, 1970.

\0[GMM72] {\sl Gallavotti, G., Martin-L\"of, A., Miracle--Sol\'e, S}:
{\it
Some problems connected with the description of coexisting phases at
low temperatures in the Ising model}, in {\sl Springer lecture notes
in Physics}, edited by A. Lenard, vol. {\bf20}, p. 159--204, Berlin
1972.

\0[GM00] {\sl Gentile, G., Mastropietro, V.}:
{\it Renormalization Group for one-dimensional fer\-mions.  A review
on
mathematical results}, preprint http://ipparco.roma1.infn.it, FM
2000-11.

\0[Gi69b] {\sl  Ginibre, J.}: {\it Dilute quantum systems}, in {\sl
Phase
transitions and critical points}, edited by C.Domb and M.S.Green,
p. 112--137, Wiley, New York, 1972.

\0[Gi71] {\sl Ginibre, J.}: {\it Some applications of functional
integration in statistical mechanics}, Statistical mechanics and
quantum field theory, ed. C. De Witt and R. Stora, p. 327--427, Gordon
Breach, 1971.

\0[La72] {\sl Lanford, O.}: {\it Entropy and equilibrium states in
classical statistical mechanics}, in {\sl Springer lecture notes in
Physics}, edited by A. Lenard, vol. {\bf20}, p. 1--113, Berlin, 1972.

\0[Le87] {\sl Lesniewski, A.:}
{\it Effective action for the Yukawa 2 quantum field Theory},
Commun. Math. Phys., {\bf 108}, 437--467, 1987.

\0[LLS99] {\sl Lebowitz, J.L., Lenci, M., Spohn, H.}:
{\it Large deviations for ideal quantum systems},
xyz.lanl.gov/abs/math-ph/9906014.

\0[MS67] {\sl Minlos, R.A., Sinai, J.G.}: {\it The phenomenon of phase
separation at low temperatures in some lattice models of a gas, I},
Math. USSR Sbornik, {\bf 2}, 335--395, 1967. And: {\it The phenomenon
of phase separation at low temperatures in some lattice models of a
gas, II}, Transactions of the Moscow Mathematical Society, {\bf19},
121--196, 1968. Both reprinted in [Si91]. The methods were exploited
in the theory of phase coexistence in the Ising model in [GMM72] which
also gives a review of such applications: see lemmas in Sec. 5 and in
the related appendix 5A, \eg eq. (5A.11).

\0[Ol88] {\sl Olla, S.}: {\it Large Deviations for Gibbs
	Random Fields}, Probability Theory and Related Fields,
	{\bf77}, 343-357, 1988.

\0[Pa85] {\sl Park, Y.M.}: {\it Quantum statistical mechanics for
	superstable interactions: Bose--Einstein statistics}, Journal
	of Statistical Physics, {\bf40}, 259--302, 1985.

\0[Ro70] {\sl Robinson, D.}: {\it Statistical mechanics of quantum
	mechanical particles with hard cores}, Communications in
	Mathematical Physics, {\bf16}, 290--309, 1970. And
	Robinson, D.: Lectrure Notes in Physics, vol. ???, 1972??.

\0[Ru69] {\sl Ruelle, D.}: {\it Statistical Mechanics}, Benjamin,
New York, 1969.

\*

\0{\it e-mail}:\\
{\tt gallavotti@roma1.infn.it\\
lebowitz@math.rutgers.edu\\
mastropi@axp.mat.uniroma2.it}
\end